\documentclass[a4paper,twoside,reqno,11pt]{article}
\usepackage{amssymb,amsmath,amsthm}
\usepackage{graphicx}
\usepackage{times}
\usepackage{color}
\leftmargini=\baselineskip
\newtheoremstyle{localthm}
	{7pt} 
	{7pt} 
	{\sl} 
	{} 
	{\bf} 
	{{\rm.}} 
	{.7em} 
	{} 

\theoremstyle{localthm}

\newtheoremstyle{localrem}
	{5pt} 
	{5pt} 
	{\rm} 
	{} 
	{\bf} 
	{{\rm.}} 
	{.7em} 
	{} 

\theoremstyle{localrem}

\def\bs{\boldsymbol}

\begin{document}
\addtolength{\baselineskip}{0.4\baselineskip}
\title{Excess deaths, baselines, Z-scores, P-scores and peaks}
\author{Laurie Davies\\
	(University of Duisburg-Essen )}
\date{\today}
\maketitle

\begin{abstract}
The recent Covid-19 epidemic has lead to comparisons of the countries suffering from it. These are based on the number of excess deaths attributed either directly or indirectly to the epidemic. Unfortunately the data on which such comparisons rely are often incomplete and unreliable. This article discusses problems of interpretation of data even when the data is largely accurate and delayed by at most two to three weeks. This applies to the Office of National Statistics in the UK,  the Statistisches Bundesamt in Germany and the Belgian statistical office Statbel. The data in the article is taken from these three sources. The number of excess deaths is defined as the number of deaths minus the baseline, the definition of which varies from country to country. In the UK it is the average number of deaths over the last five years, in Germany it is over the last four years and in Belgium over the last 11 years. This means that in all cases the individual baselines depend strongly on the timing and intensity of adverse factors such as past influenza epidemics and heat waves. This makes cross-country comparisons difficult. A baseline defined as the number the number of deaths in the absence of adverse factors can be operationalized by taking say the 10\% quantile of the number of deaths. This varies little over time and European countries within given age groups.  It therefore enables more robust and accurate comparisons of different countries. The article criticizes the use of Z-scores to measure the degree to which the countries have been hit by the corona-virus. It has been claimed that Z-scores allow direct comparisons between the countries involved. It is argued below that the opposite is the case, the Z-scores distort the actual situation as, among other weaknesses, they do not take the size of the countries into account.  Finally the problem of describing past epidemics by their timing, that is start and finish and time of the maximum, and by their effect, the height of the maximum and the total number of deaths, is considered. A similar problem occurs when analysing peaks in X-Ray diffractograms.
\end{abstract}

\tableofcontents

\quad\\
\section{Introduction} \label{ref:intro}
Data on Covid-19 deaths are often unreliable. In Germany the number of Covid-19 deaths given by the Statistisches Bundesamt is taken from the Robert Koch Institute which requires a blood test confirming the presence of the corona virus at the time of death, symptoms are not sufficient. In the UK a death certificate signed by a doctor giving Covid-19 as the cause of death would be accepted even if the the judgment of the doctor was based purely on the symptoms. People may also die due to reasons only indirectly attributable to the disease: patients may be refused treatment because of the strain put on medical services, others may not even seek hospital treatment through fear of contracting the disease. 

For these and other reasons it has been suggested that excess deaths may give a more reliable measure of the effect of the Covid-19. Excess deaths are defined as the difference between the number of present deaths and a baseline, typically defined historically as the average number of deaths over previous years. The number of deaths can be usually be accurately determined, the main problem is that  of delays. For example on the 27th May Spain increased the number of deaths by 12000 backdated to early March. To avoid such problems this paper only uses data from the UK, and in particular England and Wales, Germany and Belgium. The data is accurate and up-to-date with Germany being the laggard with delays of about three weeks.
The sources are the Office of National Statistics for the UK,  the Statistisches Bundesamt for Germany and the Belgian statistical office Statbel for Belgium. However even with such data the  there still remain problems of comparing excess deaths. The reason is the effect of past influenza epidemics on the excess deaths calculated today. In 2018 there was a major influenza epidemic in Germany. If this one year is removed from the baseline the number of excess deaths for the weeks 11-21 increases from 7602 to 11639. It is as though the influenza epidemic of 2018 masks about 4000 excess deaths. The problem can largely be avoided by taking the baseline to be the 10\% quantile rather than the mean of deaths of  previous years.

One widely used statistic for comparing excess deaths across countries is the Z-score. In particular it is the main means of comparison used by Euromomo\\
https://www.euromomo.eu/\\
Under the ``Objectives'' one reads
{\it
\begin{quotation}
The aim of the EuroMOMO project (European monitoring of excess mortality for public health action) is to operate coordinated timely mortality monitoring and analyses in as many European countries as possible, using a standardized approach to ensure that signals are comparable between countries.
\end{quotation}
}
and under the heading ``What is a z-score?''
{\it
\begin{quotation}
Z-scores are used to standardize series and enable comparison mortality pattern between different populations or between different time periods. The standard deviation is the unit of measurement of the Z-score. It allows comparison of observations from different normal distributions.
\end{quotation}
}
Unfortunately comparison of Z-scores without knowledge of the population sizes is impossible. Such comparisons can distort the effects of an epidemic in different countries by a large factor. We show why this is so and why the use of Z-scores should be abandoned. It can be replaced with the much more intuitive concept of a P-score.

Epidemics exhibit themselves in the data as peaks. Quantifying epidemics is then the problems of localizing and evaluating the sizes of peaks. In the context of epidemics we take this to mean the quantification of an epidemic through its start and finish, its peak and the total number of deaths it caused. A similar problem occurs in the X-Ray diffractograms and we adopt the approach taken in \cite{DAGAMEMEMI08}.

\section{Excess deaths, the baseline and the age distribution}
The number of excess deaths in any week is defined as the number of deaths in that week minus the baseline for that week.  Little attention has been given to the baseline, in particular there seem to be no comparisons of the baselines of different countries. To show how large the baseline effect can be we consider data for England and Wales, Germany and Belgium.

\subsection{Historical baselines}
\begin{figure}[t]
  \centering
\includegraphics[width=.8\textwidth,height=150px]{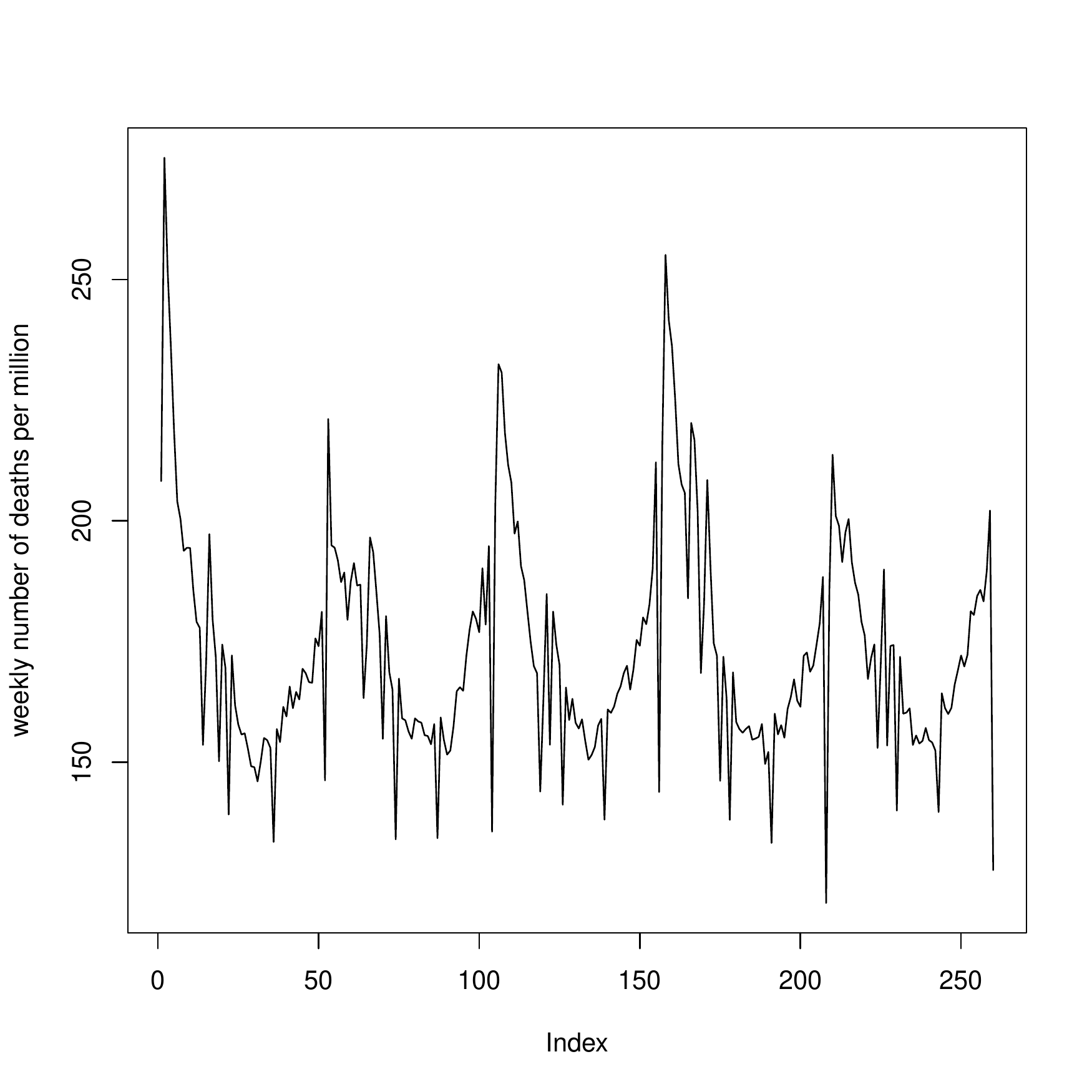}
\includegraphics[width=.8\textwidth,height=150px]{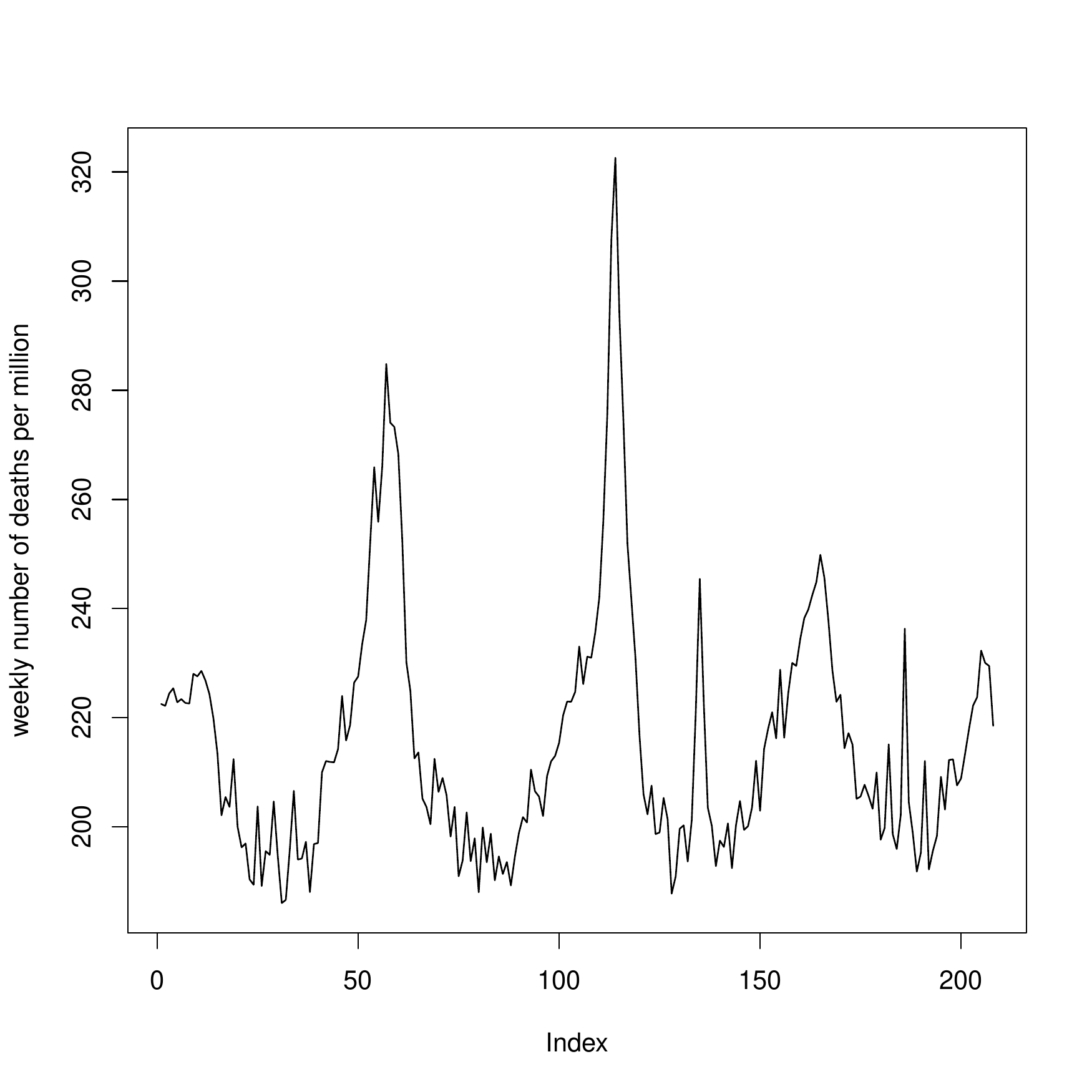}
\includegraphics[width=.8\textwidth,height=150px]{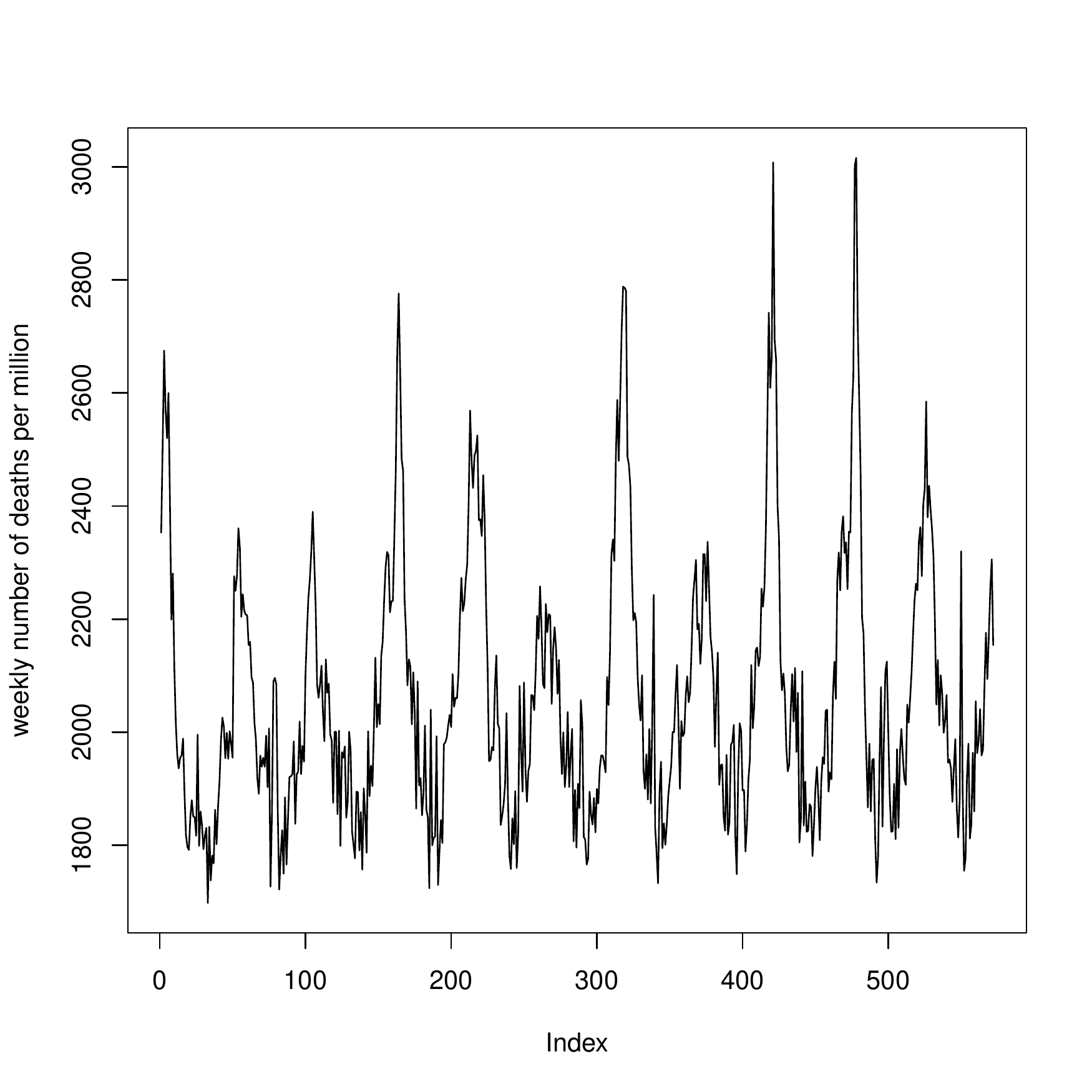}
\caption{Top panel:weekly number of deaths per million  for England and Wales 2015-2019.  Centre panel: weekly number of deaths per million for Germany  2016-2019. Bottom panel: weekly number of deaths per million for Belgium  2009-2019}
\label{fig:weekly_deaths_ew_de}
\end{figure}

\begin{figure}[t]
  \centering
\includegraphics[width=.8\textwidth,height=150px]{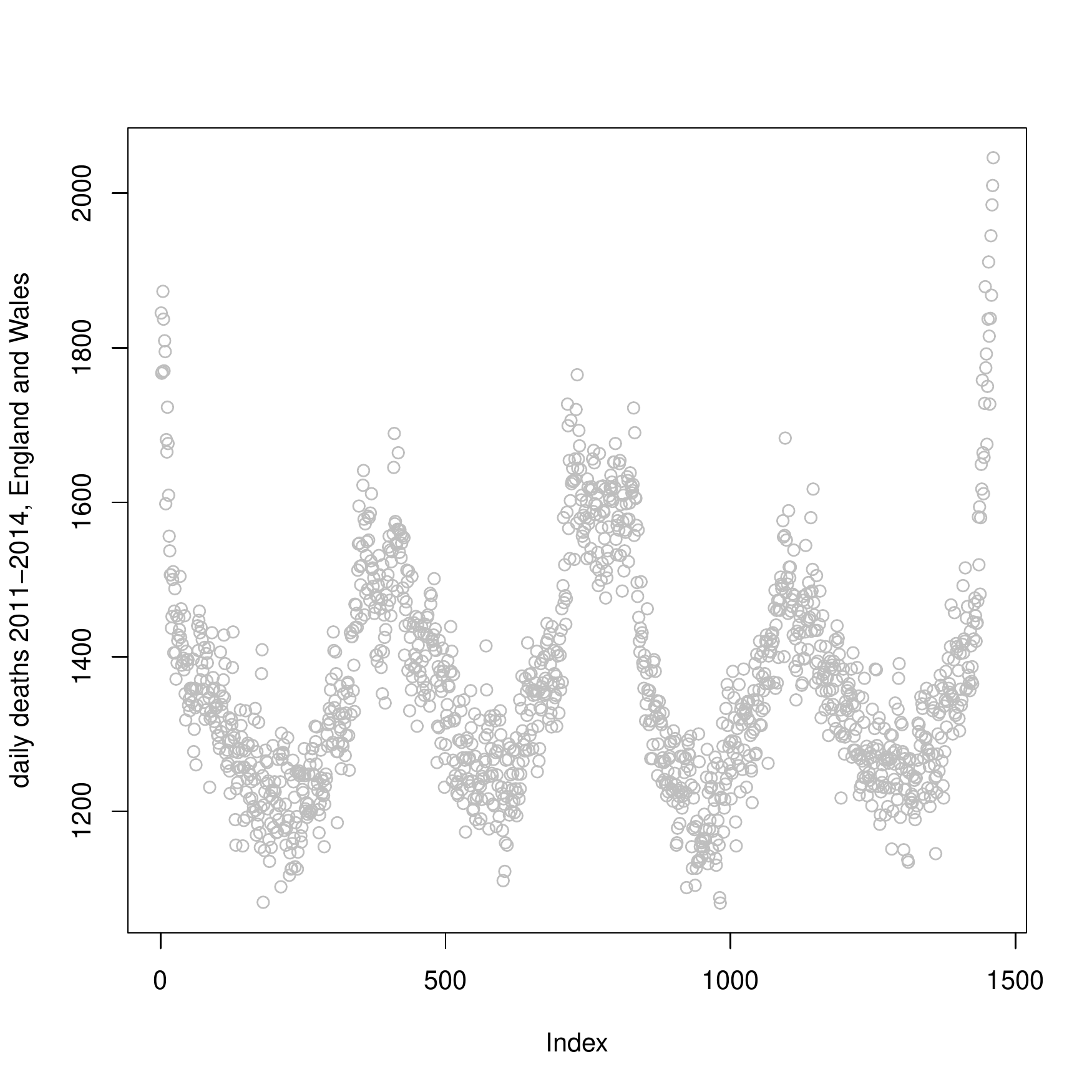}
\caption{The number of daily deaths England and Wales 2011-2014.}
\label{fig:daily_deaths_ew}
\end{figure}

The top panel of Figure~\ref{fig:weekly_deaths_ew_de} shows the number of weekly deaths per million of population for England and Wales for the years 2015-2019. The centre panel shows the same for Germany for the years 2016-2019 and the bottom panel the same for Belgium for the years 2009-2019. We note that there are large drops in the number of weekly deaths in England and Wales for weeks which include the Spring, Summer and Christmas Bank Holidays. These disappear if daily data is considered as shown in Figure~\ref{fig:daily_deaths_ew} which plots the daily deaths from 2011-2014. All sets of data show influenza epidemics of varying intensity and length. The German and Belgian data also show deaths due to summer heat waves.

The baselines derived from the two three data sets are shown in Figure~\ref{fig:hist_bsln_de_ew}, that for and England and Wales by (o), for Germany by (+)  and Belgium by (*).  The three outliers in the England-Wales baseline are the Spring, Summer and Christmas Public holidays mentioned above. The influenza epidemics are visible in all  baselines with the peak of the German epidemics occurring about six week later than those of England and Wales. The increase in deaths due to heat waves is also visible in the German baseline and to a lesser extent in the Belgium baseline.

\subsection{Age distribution}
The main difference between the the baselines in the top panel is that the German one is clearly higher than than the other two. This is due to the difference in age distributions. England and Wales have a median age of about 40, Germany one of 47 and Belgium one of 42. There are about 11 million people over the age of 65 in England and Wales, about 17.5 million in Germany and 2.2 million in Belgium. The respective populations are approximately 59, 83 and 11.5 million so that the proportion of people over the age of 65 are respectively 18.6\%, 21.1\% and 19\%. The effect can be eliminated to some extent by restricting attention to a particular age groups and calculating a baseline for each age group. We do this for the age groups over 65 and under 65 shown in the centre and bottom panels. The England and Wales baseline for age group under 65 is somewhat erratic.

\begin{figure}[t]
  \centering
\includegraphics[width=.8\textwidth,height=150px]{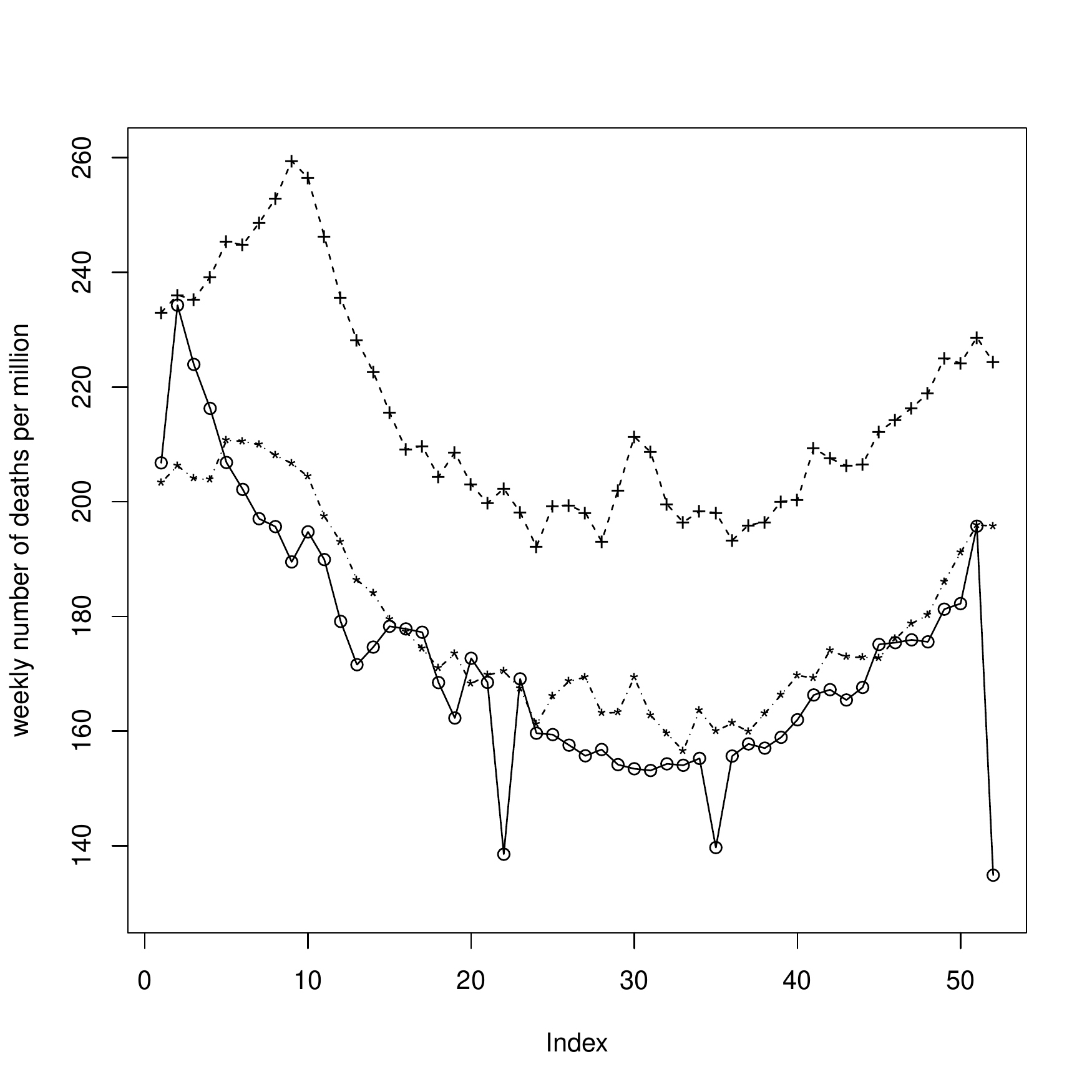}
\includegraphics[width=.8\textwidth,height=150px]{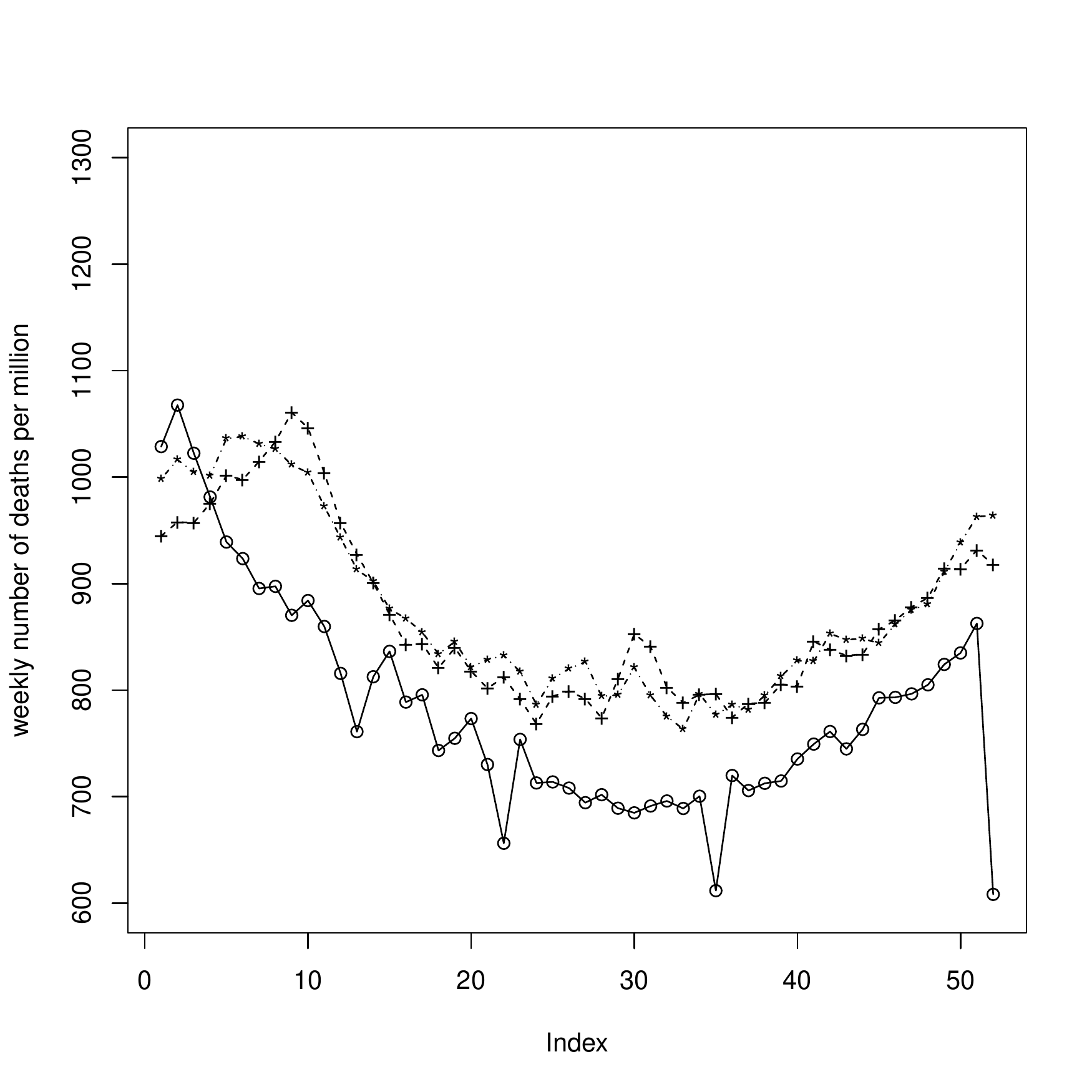}
\includegraphics[width=.8\textwidth,height=150px]{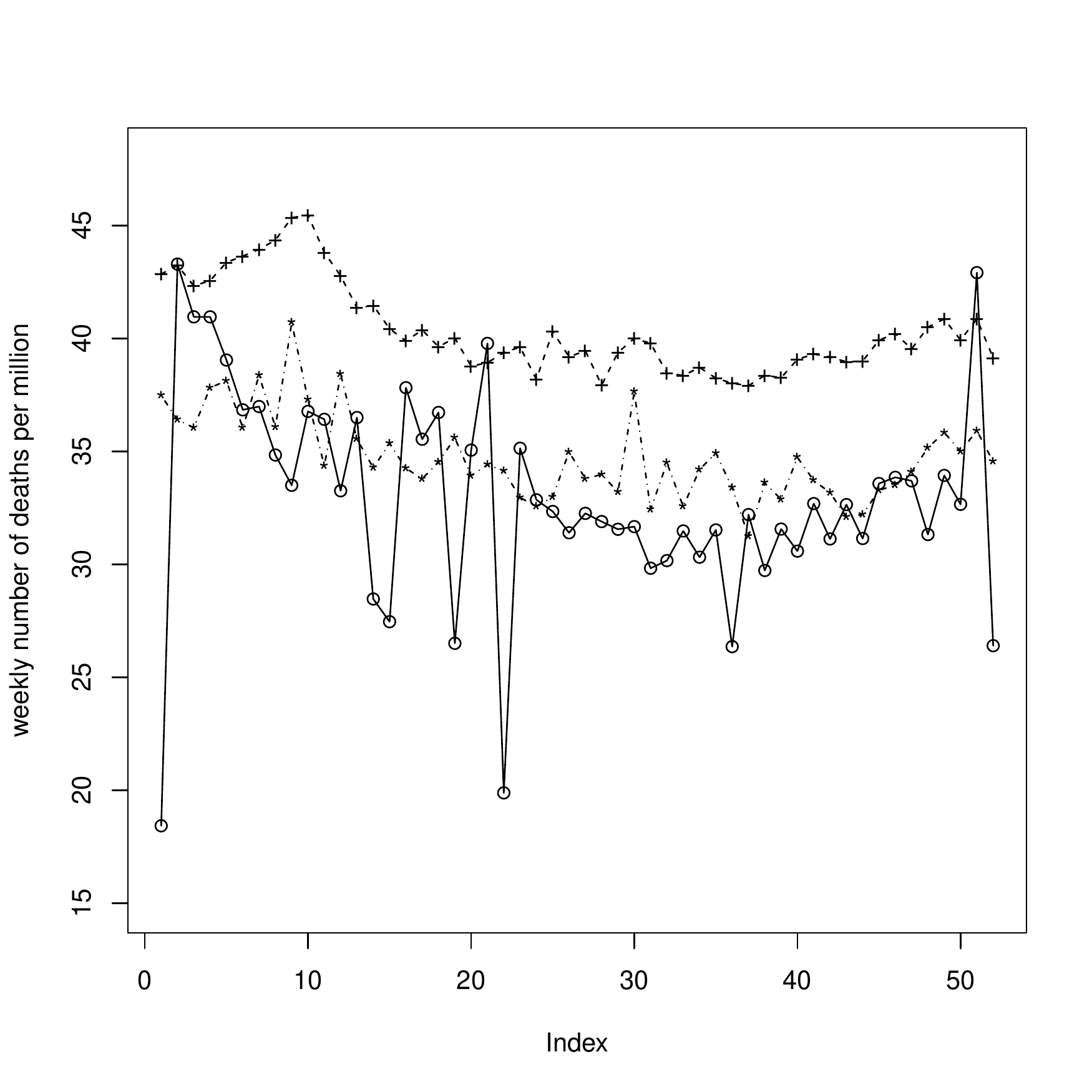}
\caption{The historical baselines, number of deaths per week per million of population for  England and Wales (solid), Germany (dashed) and Belgium (dotdash). Top panel; all ages, centre; over 65, bottom; under 65. }
\label{fig:hist_bsln_de_ew}
\end{figure}

\begin{figure}[t]
  \centering
\includegraphics[width=.8\textwidth,height=150px]{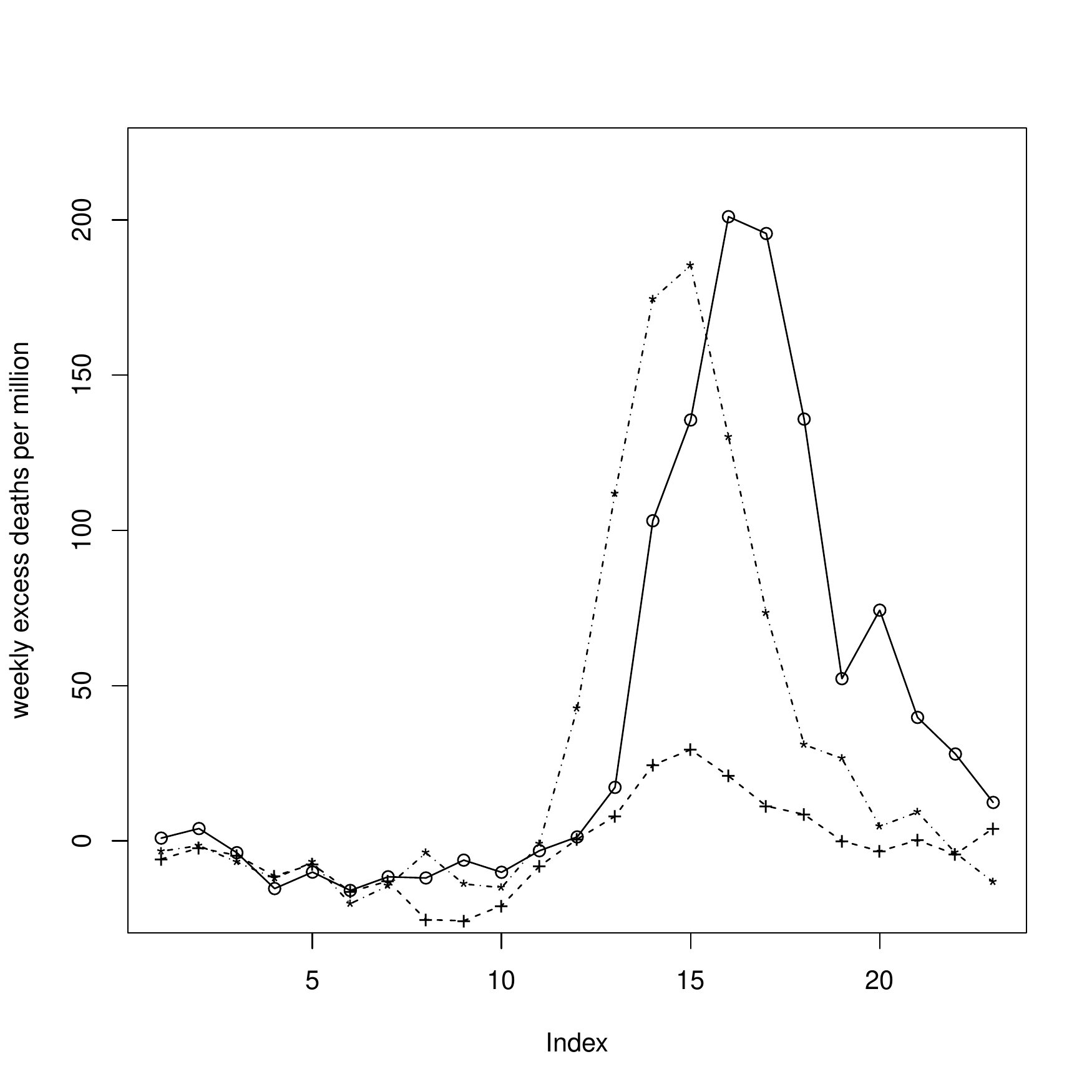}
\includegraphics[width=.8\textwidth,height=150px]{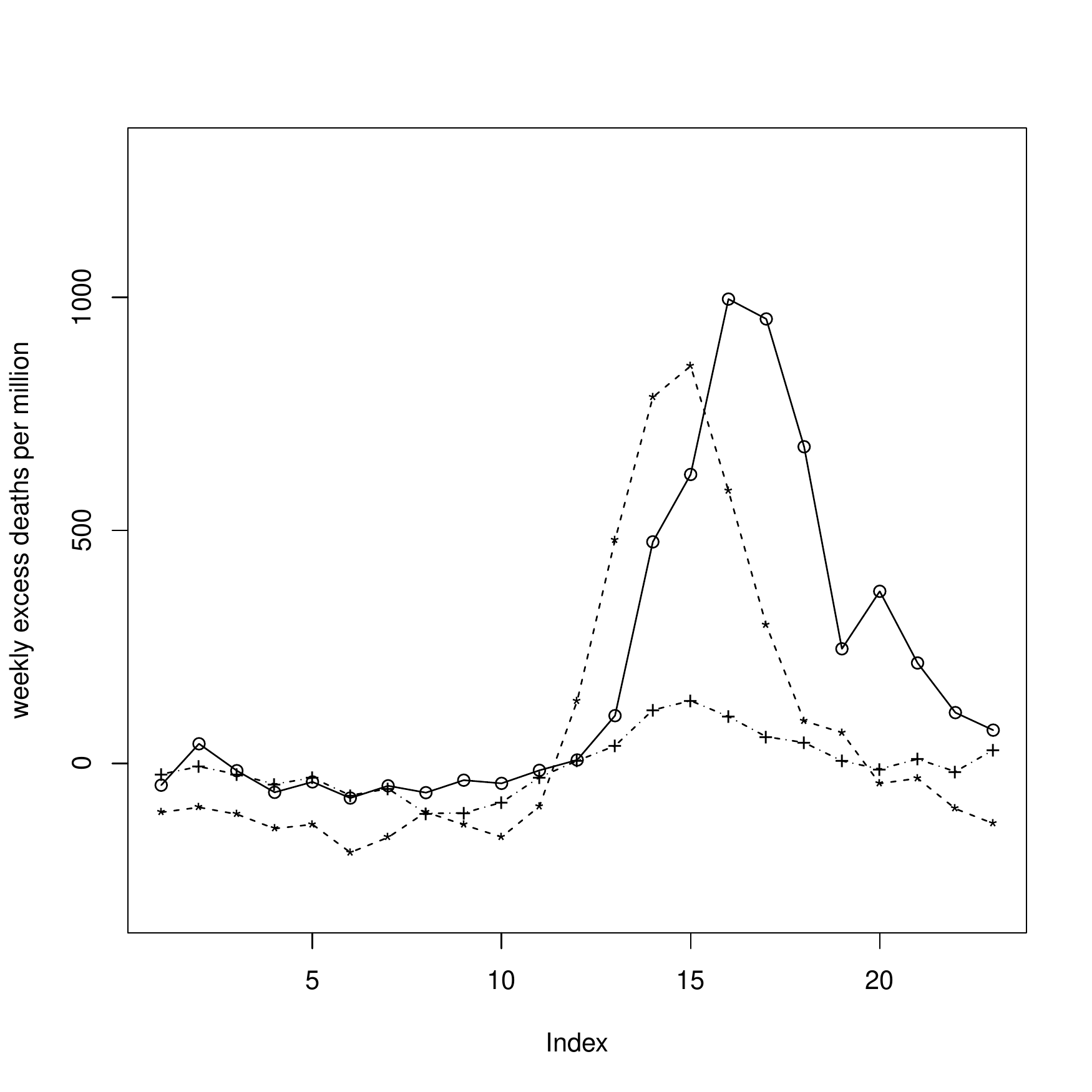}
\includegraphics[width=.8\textwidth,height=150px]{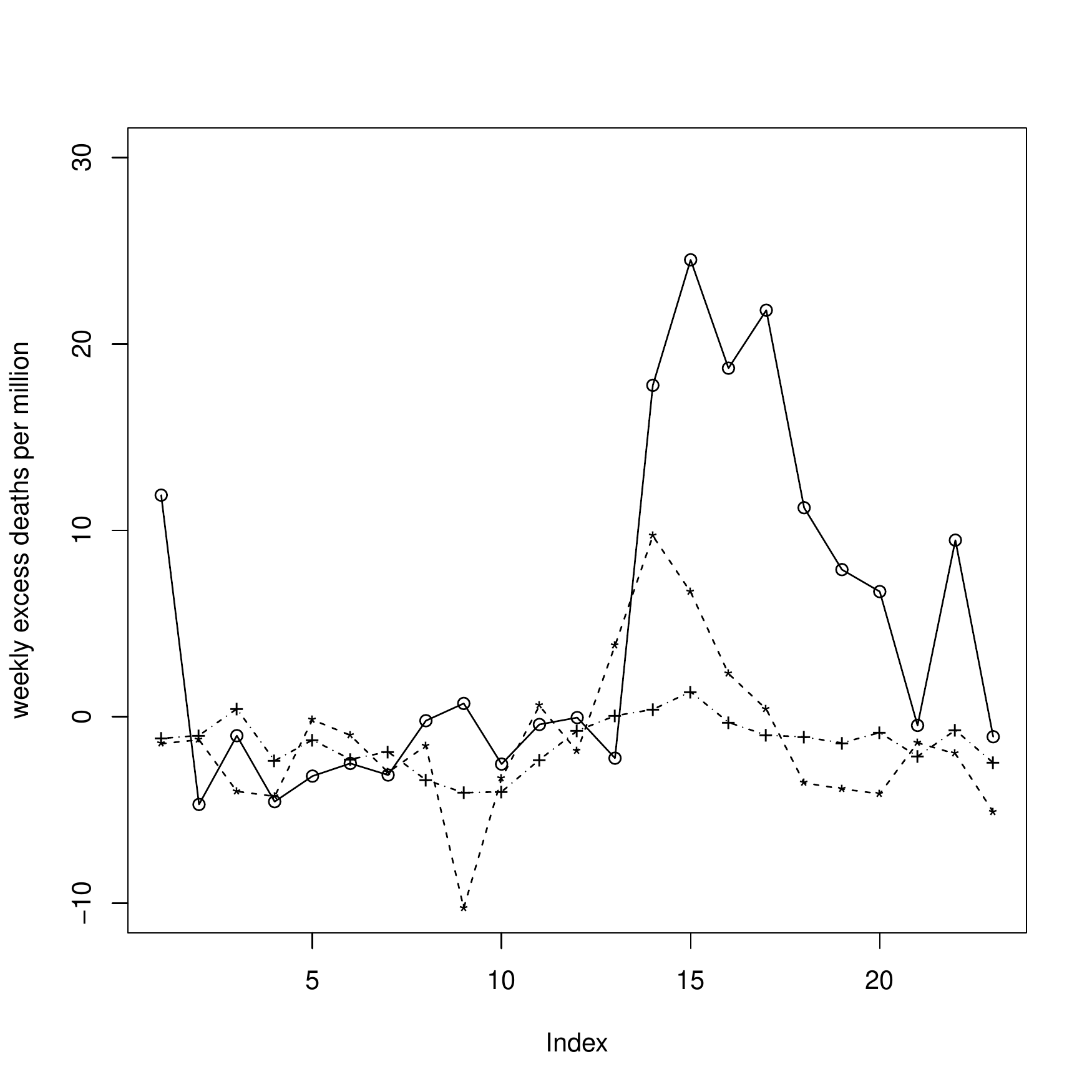}
\caption{The weekly excess deaths for  England and Wales (o), for Germany (+), and Belgium (*) for the first 23 weeks of 2020 using the historical baselines of Figure~\ref{fig:hist_bsln_de_ew}: top panel, all ages; centre, over 65; bottom, under 65.}
\label{fig:de_ew_excess_deaths}
\end{figure}

The top panel of Figure~\ref{fig:de_ew_excess_deaths} shows the number of excess deaths for the first 21 weeks of 2020 derived from the baselines of  Figure~\ref{fig:hist_bsln_de_ew} for England and Wales (o), for Germany (+) and Belgium (*). The centre and bottom panels show the same for the ages groups over 65 and under 65 respectively. 

Figure~\ref{fig:weekly_deaths_ew_de} and the baselines of  Figure~\ref{fig:hist_bsln_de_ew} show that the influenza epidemics in Germany are more variable than those in England and Wales and that they peak at about week nine against week two in England and Wales. This suggests that the German excess death numbers are strongly influenced by the choice of the historical baseline. The German baseline of the top panel of Figure~\ref{fig:hist_bsln_de_ew} results in 7602 excess deaths for the weeks 11:23. If the year 2018 is removed from the baseline the number becomes 11639. Thus the influenza epidemic of 2018 reduces the number of excess deaths by about 4000. One can also look at the timing of the Covid-19 epidemic. If we shift the data for weeks 11-23 to 10-22 , that is one week earlier,  the number of excess deaths would have been 2760, if it had occurred one week later the number would have been 12094. The corresponding numbers for England and Wales are 57088 and 63088 and for for Belgium  8446 and 9289.

The variability of the historical baselines is ignored in all discussions of the Covid-19 epidemic. We propose another definition of the baseline, namely the number of deaths in the absence of any detrimental causes such as influenza epidemics and heat waves. Other causes could be listed from road deaths to the absence of health care but in the context of Covid-19 we consider only influenza  and heat waves. Based on this we suggest the median of the 10th quantile of the weekly number of deaths over previous years. The values of the 10\% quantiles are given in Table~\ref{tab:10_quantile}. They are remarkably constant over the years. The median of the four  10\% quantiles all ages for Germany is 192 which results in 363 excess deaths per million of the population for the weeks 11-21. If the individual 10\% quantiles of the four years are used the number of excess deaths per million of the population ranges from 330 to 396. I
\begin{table}[h]
\begin{center}
\begin{tabular}{rrrrr}
&2016&2017&2018&2019\\
\hline
ew0+&152& 150& 149& 152\\
de0+&189&191& 193& 195\\
be0+&158& 158& 158& 159\\
ew65+&679& 670& 670& 686\\
de65+&755& 765& 774& 792\\
be65+&684&688& 687& 704\\
ew064-&29& 28& 29& 29\\
de064-&38& 37& 37&37\\
be064-&34&34&32&32\\
\end{tabular}
\caption{10\% quantiles of weekly excess deaths per million of the population for the years 2016-2019 for England and Wales, Germany  and Belgium for the age groups 0+, 65+ and 65-}
\label{tab:10_quantile}
\end{center}
\end{table}

Daily data are available for England and Wales for 1970-2014. The 10\% quantile is 1289 daily deaths. Multiplying by seven and dividing by 59 to give number of weekly deaths per million of the population results in the value  152.9 agreeing very well with Table~\ref{tab:10_quantile}. The same procedure fo Belgium using the daily deaths from 2009-2019  results in 153.4 somewhat lower than the values in Table~\ref{tab:10_quantile}.

Table~\ref{tab:deaths_per_million_2020} gives the number of excess deaths per million of population for the three countries and the three age groups using the baselines of Table~\ref{tab:10_quantile}. In Germany for the weeks 11-23 there were 378 excess deaths per million for the whole population based on the quantile baseline. The number of confirmed Covid-19 deaths is about 9000 or 108 per million of the population. This leaves 270  excess deaths per million of the population. How many of these may be put down to the Covid-19 epidemic depends on the intensity o the influenza epidemic of the winter 2019-2020 and how quickly  it declines.  
\begin{table}[h]
\begin{center}
\begin{tabular}{crrrrrrr}
&\multicolumn{2}{c}{0+}&\multicolumn{2}{c}{65+}&\multicolumn{2}{c}{64-}\\
weeks&1:10&11:23&1:10&11:23&1:10&11:23\\
\hline\\
ew&402&1161&1964&5605&36&132\\
&-428&993&-388&4830&-9&114\\
de&398&378&1391&1241&28&10\\
&-132&92&-551&476&-21&-11\\                 
be&382&1019&1965&5263&30&45\\
&-98&771&-1326&2895&-30&2\\
\end{tabular}
\caption{Number of weekly excess deaths per million in 2020 for England and Wales, Germany,  and Belgium for the age groups 0+, 65+ and 64-. The first lines are based on the quantile baseline, the second on the historical baseline.}
\label{tab:deaths_per_million_2020}
\end{center}
\end{table}

\section{Z-scores and P-scores}\label{sec:Z-r_score}
\subsection{Z-scores}
Given data $\bs{X}_n=(X_1,\ldots,X_n)$ with mean $\bar{\bs{X}}_n$ we model the data as i.i.d. $N(\mu+\delta,\sigma^2)$  with known $\mu$ and $\sigma$. We require a confidence interval for $\delta$. The Z-score is
\[Z_{sc}=\sqrt{n}(\bar{\bs{X}}_n-\mu)/\sigma\]
and as $\bar{\bs{X}}_n=\mu+\delta+Z\sigma/\sqrt{n}$ with $Z=N(0,1)$ we have
\[Z_{sc}=\sqrt{n}(\mu+\delta+Z\sigma/\sqrt{n}-\mu)/\sigma=\sqrt{n}\delta/\sigma+Z\]
which gives
\[\delta=\sigma Z_{sc}/\sqrt{n}-\sigma Z/\sqrt{n}.\]
Thus with probability 0.95
\[\vert \delta-\sigma Z_{sc}/\sqrt{n}\vert \le 1.96\sigma/\sqrt{n}.\]
One notes that the Z-score is divided by $\sqrt{n}$.

We now consider the case where $\bs{X}$ satisfies a Poisson distribution with mean $\lambda(1+\delta)$ with known $\lambda$. Again a confidence interval is required for $\delta$. The Z-score is
\[Z_{sc}=(\bs{X}-\lambda)/\sqrt{\lambda}.\] 
Now 
\[(\bs{X}-\lambda(1+\delta))/\sqrt{\lambda(1+\delta)}\approx N(0,1)\]
for large $\lambda$. For example for $\lambda=300$ the Kolmogorov distance between distribution functions of $(\bs{X}-\lambda)/\sqrt{\lambda}$ and $N(0,1)$ is less than 0.01. Thus we may write
\[\bs{X}=\lambda(1+\delta))+Z\sqrt{\lambda(1+\delta)}\]
with $Z\approx N(0,1)$ and hence
\[Z_{sc}=(\lambda\delta +Z\sqrt{\lambda(1+\delta)}\,)/\sqrt{\lambda}=\sqrt{\lambda}\delta+Z\sqrt{1+\delta}\]
so that
\[\delta-Z_{sc}/\sqrt{\lambda}=Z\sqrt{(1+\delta)/\lambda\,}\]
giving a 95\% confidence interval for $\delta$
\[\left\{\delta: \vert \delta-Z_{sc}/\sqrt{\lambda}\vert\le 1.96\sqrt{(1+\delta)/\lambda}\,\right\}\]
If $\lambda$ is the number of weekly deaths in a large European population of size $n$ then $\lambda\approx0.01n/52=0.00019n$ so the confidence interval becomes
\[\left\{\delta: \vert \delta-72Z_{sc}/\sqrt{n}\vert\le 141\sqrt{(1+\delta)/n}\,\right\}\]
If there is over dispersion, that is the variance of $\bs{X}$ is $\sigma^2\lambda$ for some $\sigma$ the confidence interval becomes
\[\left\{\delta: \vert \delta-72Z_{sc}/\sqrt{n}\vert\le 141\sigma \sqrt{(1+\delta)/n}\,\right\}\]
Without knowledge of $n$ and $\sigma$ no information about $\delta$ can be obtained simply from the Z-score. The Z-scores published by Euromomo are more complicated than this although exactly how they are calculated is not known. Whatever the case they are simply misleading and consequently worthless. 

\subsection{P-scores}
\begin{figure}[t]
  \centering
\includegraphics[width=.8\textwidth,height=150px]{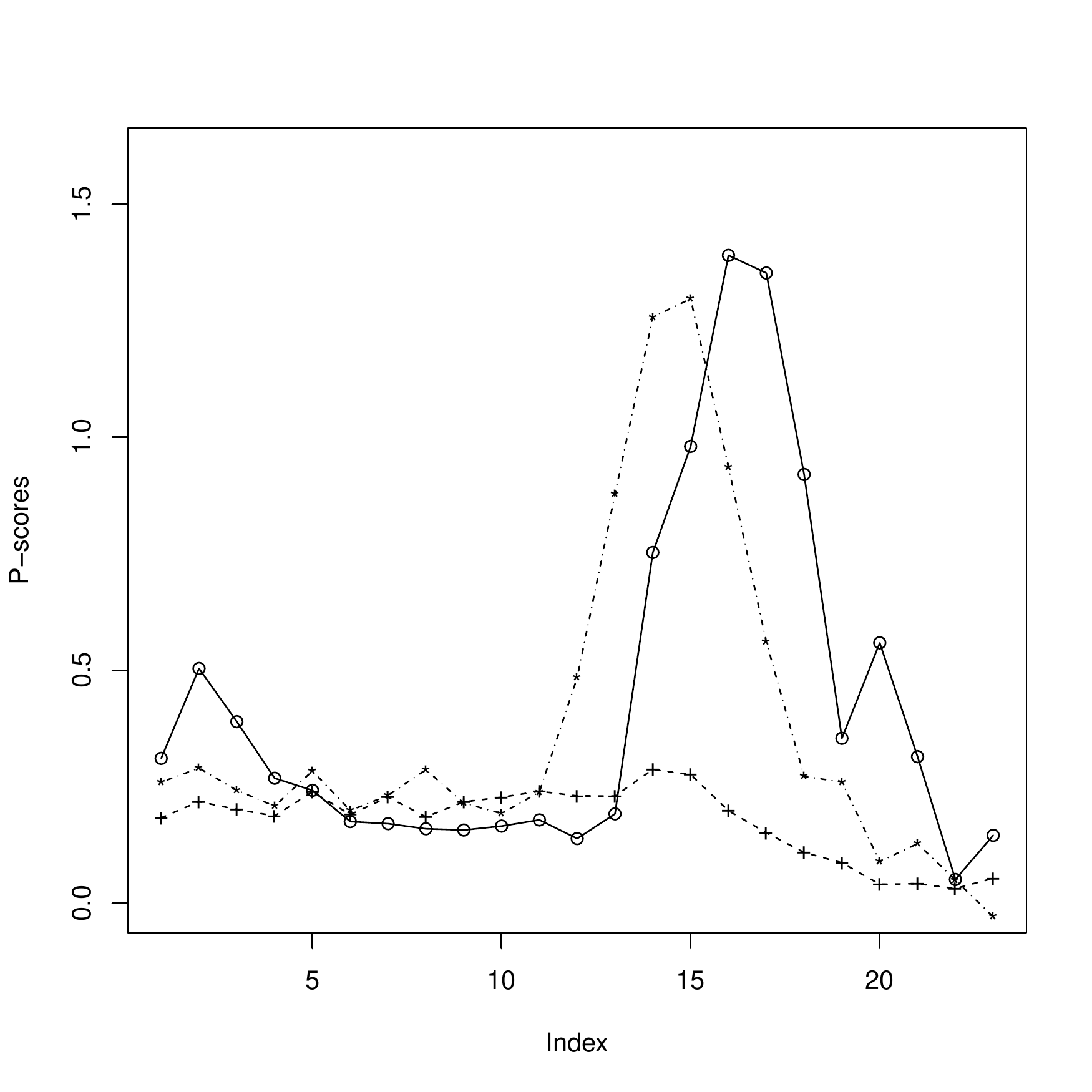}
\includegraphics[width=.8\textwidth,height=150px]{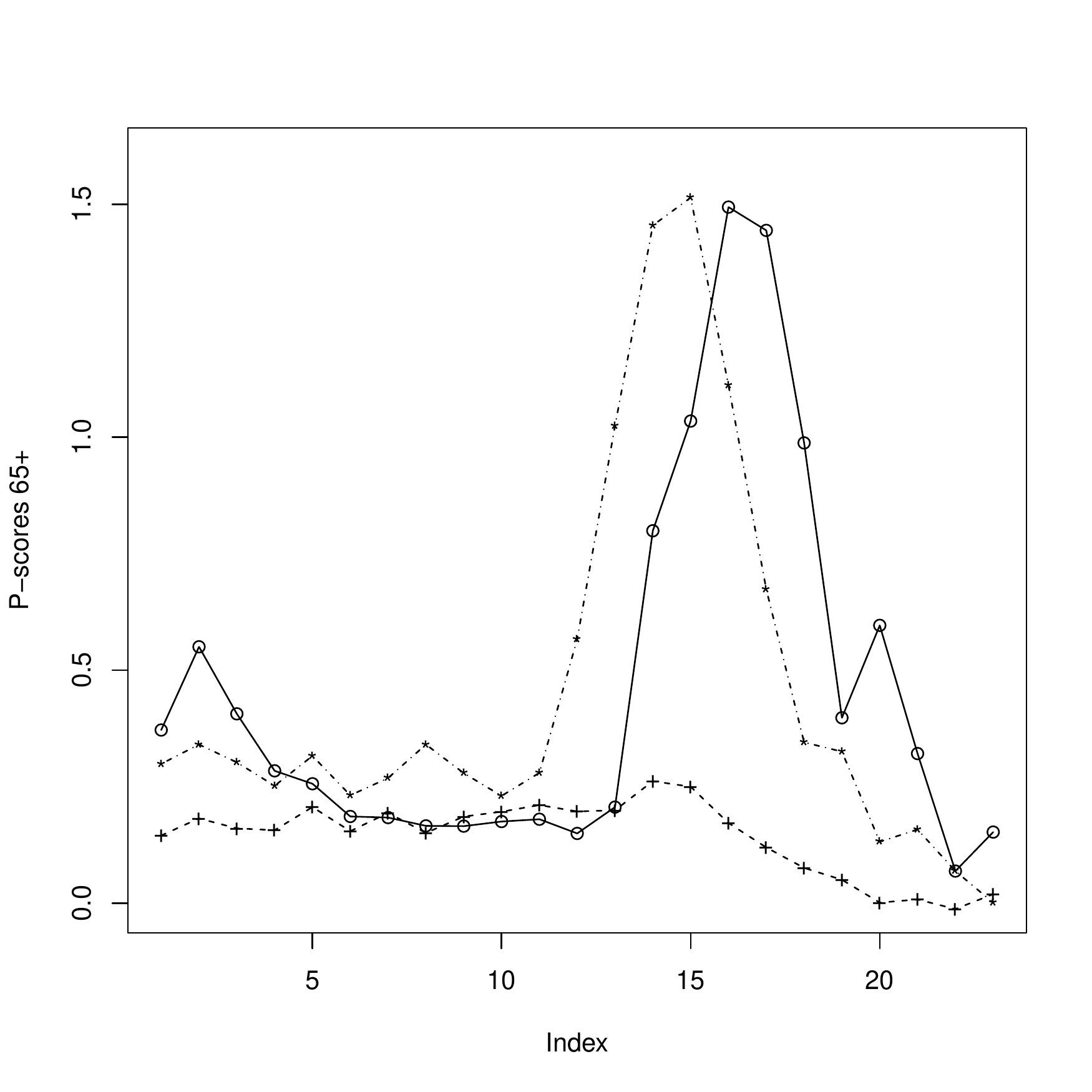}
\includegraphics[width=.8\textwidth,height=150px]{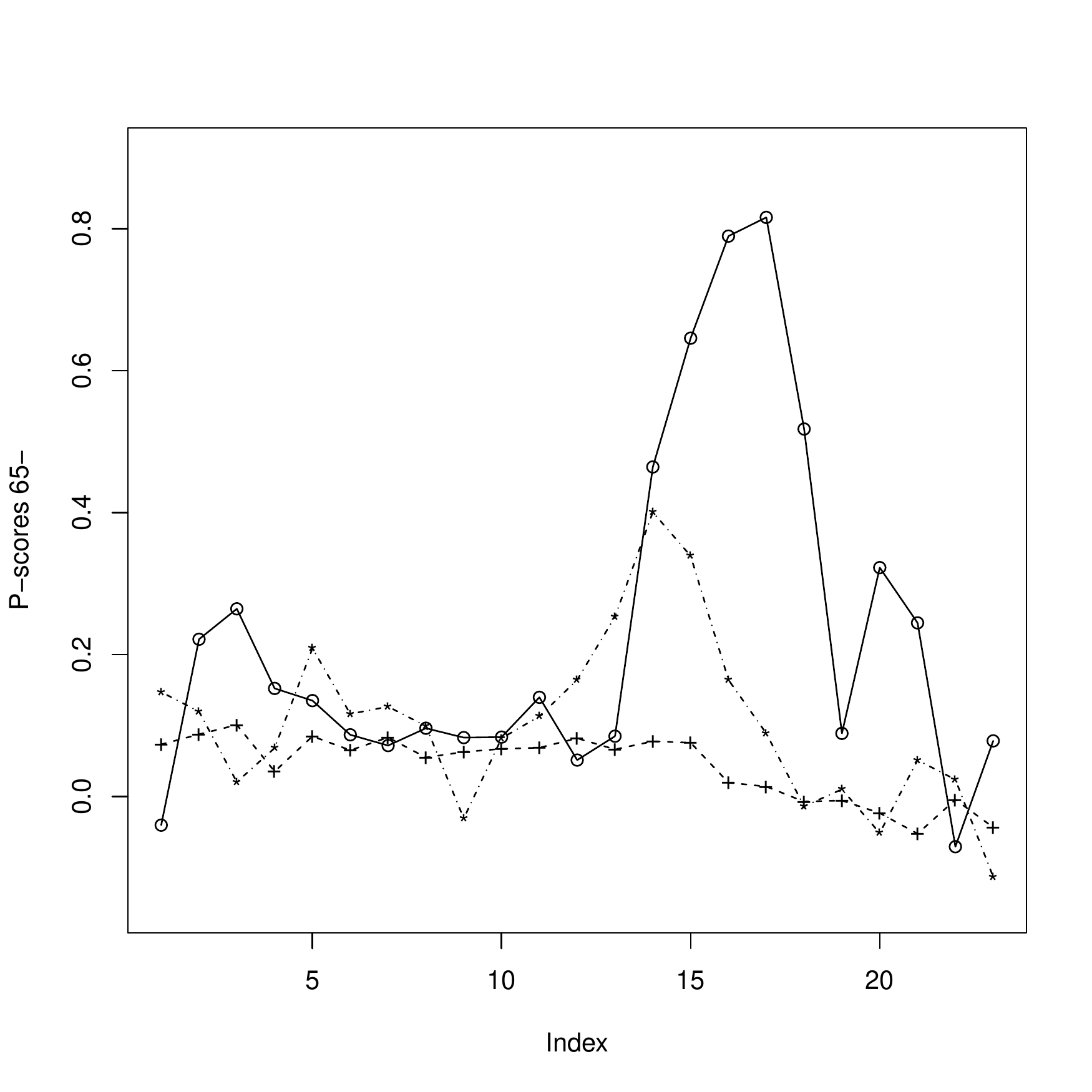}
\caption{The P-scores for England and Wales  (o),  Germany (+) and Belgium (*). Top panel: all age groups; Centre panel: over 65; Bottom panel: under 65.}
\label{fig:R_de_ew_bel}
\end{figure}

We define the P-score for the {\it i}th week as $(x(i)-bsln(i)/bslin(i)$ where $x(i)$ is the number of deaths and $bsln(i)$ the value of the baseline for that week (see \cite{ARMU20}). In contrast  to the Z-scores the P-scores do not depend on the size of the population. The P-scores for the three countries and three age groups considered here are shown in Figure~\ref{fig:R_de_ew_bel}.

\section{Models}\label{sec:model}
In his book \cite{HUB11} Peter Huber points out that the word `model' has `a bewilderingly wide semantic range' but with a common feature
\begin{quotation}
a model is a representation of the essential aspects of some real thing in an idealized, exemplary form, ignoring the inessential ones.
\end{quotation}
EuroMOMO aims 
\begin{quotation}
to detect and measure excess deaths related to seasonal influenza, pandemics and other public health threats. 
\end{quotation}
Given this it is strange that EuroMOMO should choose the Poisson model which, with a single parameter, is too simple to enable EuroMOMO to accomplish its aims. Because of this EuroMOMO has to perform several operations on the original data before calculating the $Z$-scores: 
\begin{quotation}
Z-score(s) are computed on the de-trended and de-seasonalized series, after a 2/3 powers transformation according to the method described in Farrington et al. 1996. This enables the computation of Z-scores for series that are originally Poisson distributed.
\end{quotation}
 The aims and the calculation of the $Z$-scores seem to contradict each other. If you want to measure excess deaths including seasonal data it seems contra-productive to de-trend and de-seasonalize. The 2/3 powers transformations of \cite{FAANBECA96} are to take into account the asymmetry of the Poisson distribution when calculating $Z$-values. The smallest value of $\lambda$ for the 24 countries considered by EuroMOMO is about 350 for Northern Ireland. The maximum difference between the cumulative distribution functions of the $\text{Po}(350)$ and its normal approximation is about 0.014 so it hardly seems worthwhile making this correction. Moreover you can calculate the $Z$-score exactly without the normal approximation. Suppose the data are $x$ and the model is  $\text{Po}(\lambda)$. The P-value for $x$ is $p=\bs{P}_{\lambda}(X \ge x)$ which can be calculated using any statistical software. The corresponding $Z$-score for $x$ is $z=\Phi^{-1}(1-p)$ where $\Phi$ is the distribution $N(0,1)$ distribution function. 

Even all this is not sufficient. There is over dispersion which must be estimated, also local trends, restriction to spring and autumn months: see 
\begin{verbatim}
https://www.euromomo.eu/how-it-works/methods/
\end{verbatim} 

\begin{figure}
  \centering
\includegraphics[width=.8\textwidth,height=150px]{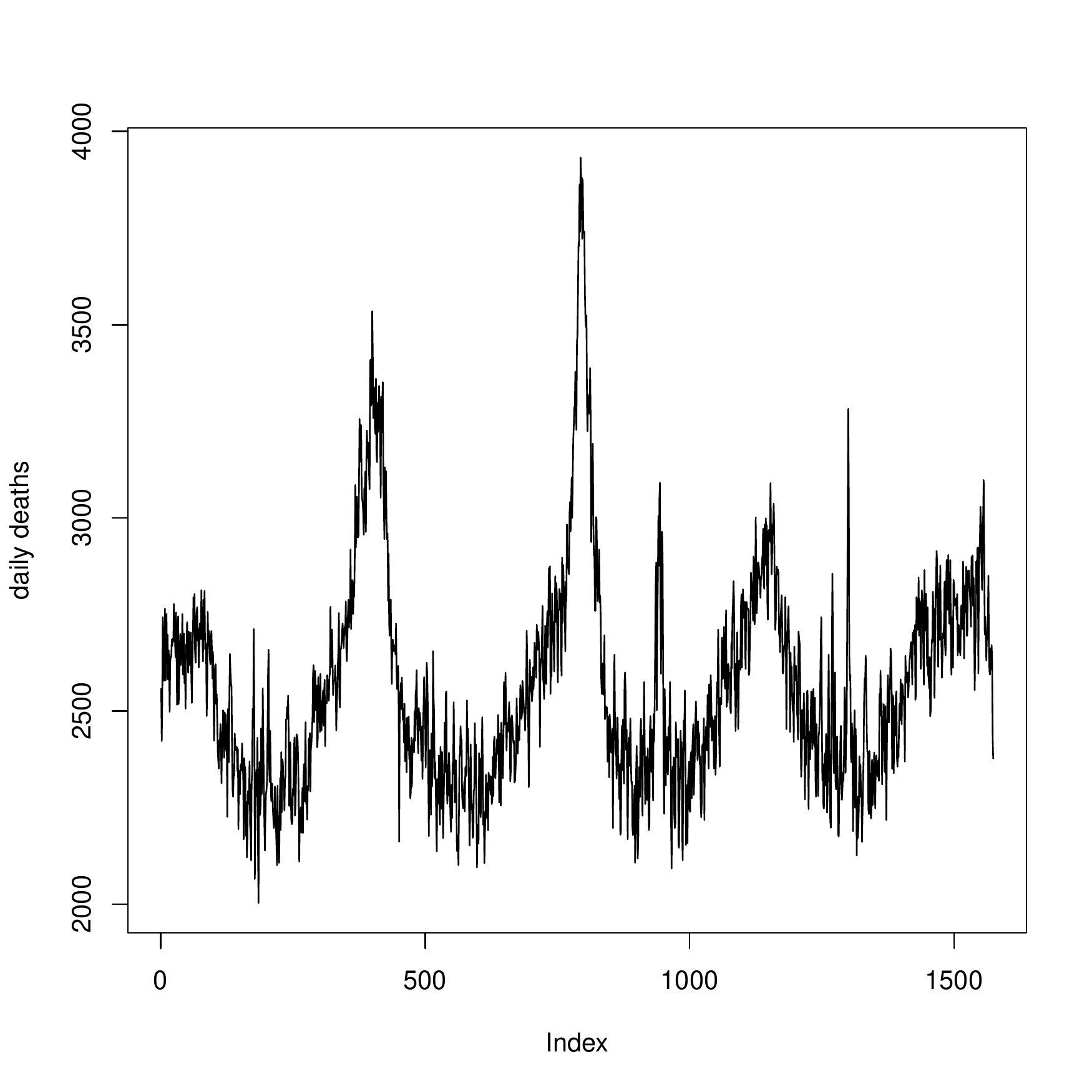}
\includegraphics[width=.8\textwidth,height=150px]{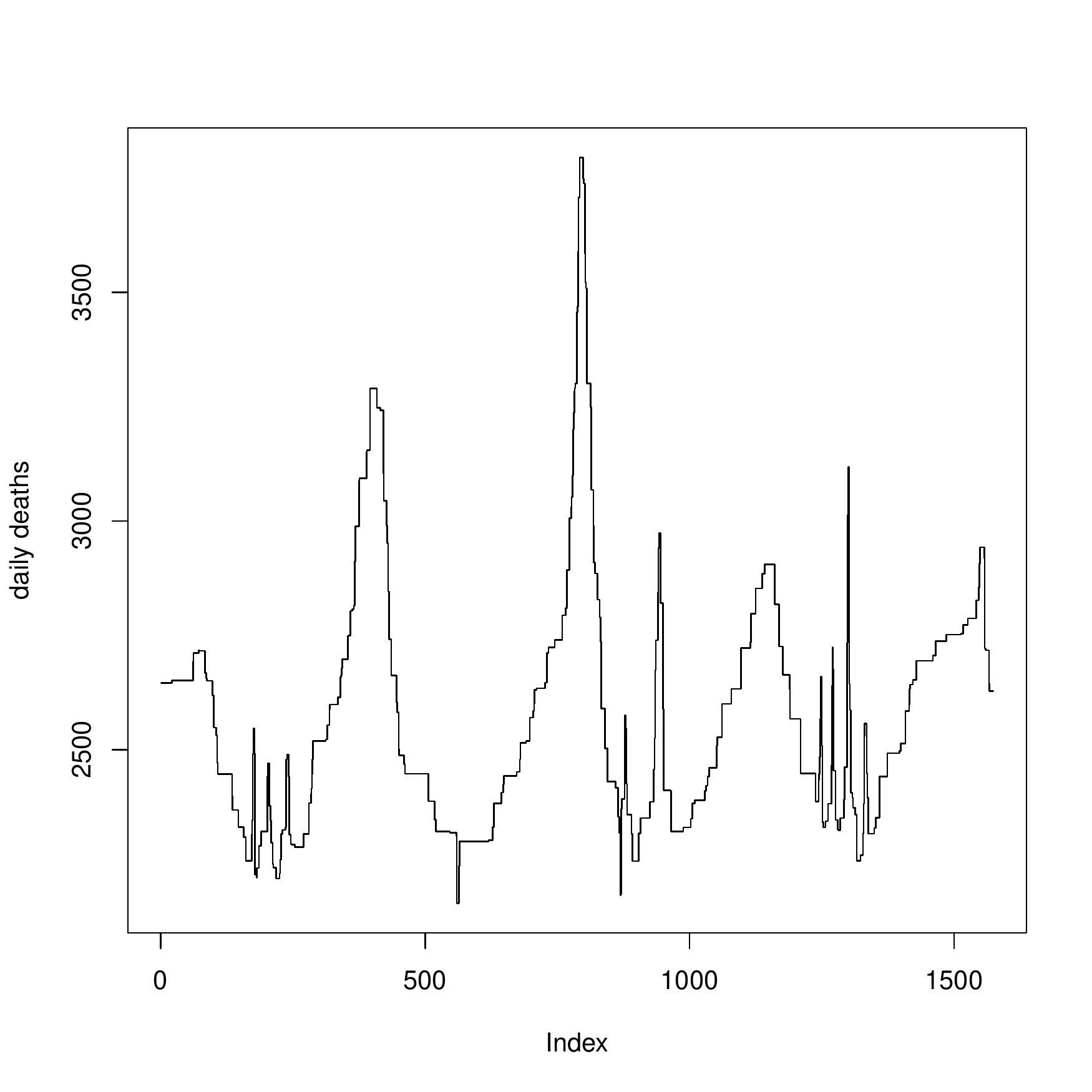}
\includegraphics[width=.8\textwidth,height=150px]{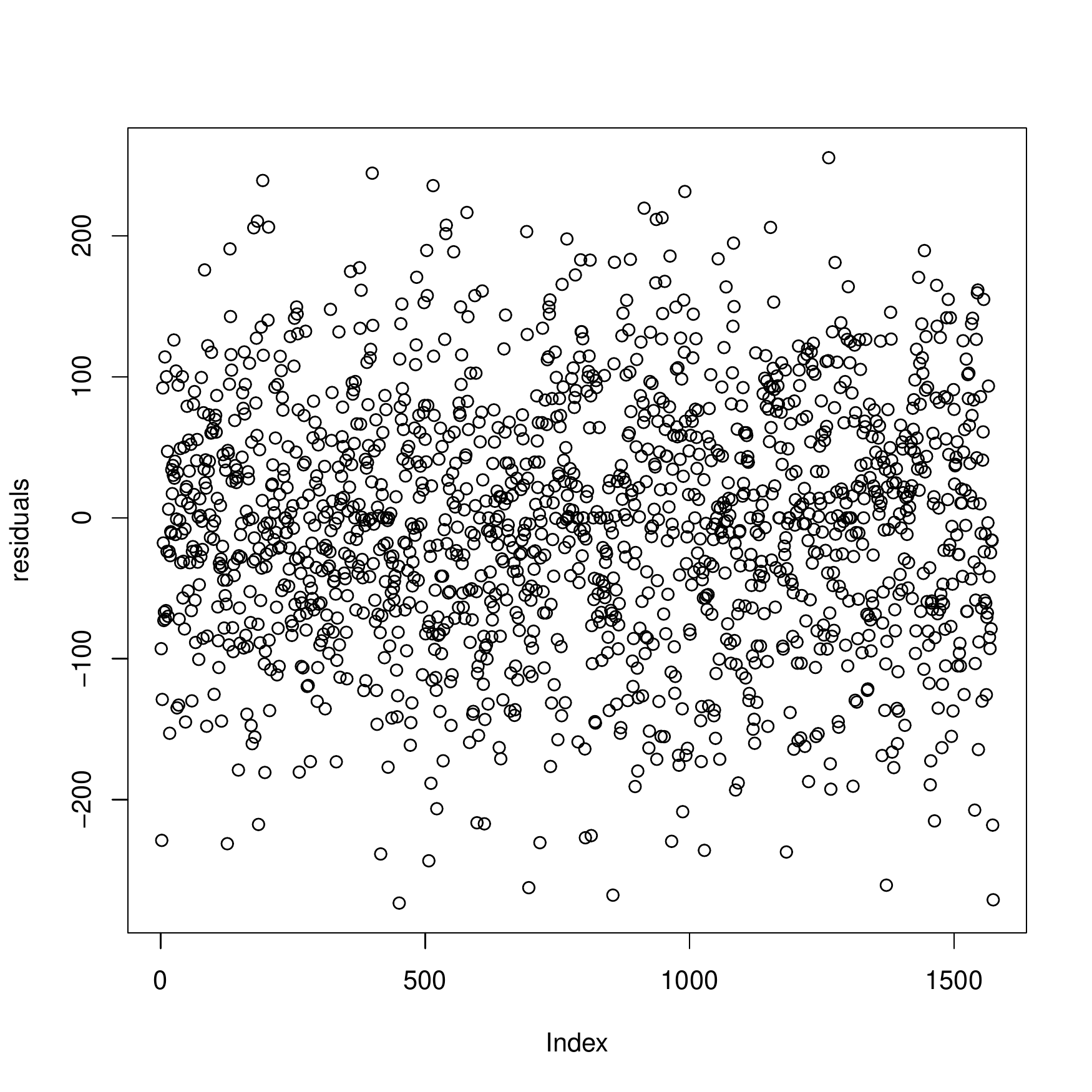}
\caption{Top panel: the number of daily deaths in Germany from  01.01.2016-26.04.2020. Center panel: the taut string regression function for this data. Bottom panel: the residuals from the taut string }
\label{fig:mort}
\end{figure}

Excess deaths due to influenza, pandemics, heat-waves and other public health threats manifest themselves in the data as peaks in the data: see Figure~\ref{fig:mort} which gives the number of daily deaths in Germany from 01.01.2016-26.04.2020. A sharp peak in July 2019 is clearly visible caused by temperatures of around 40C. Thus the aims of EuroMOMO stated above can be translated into the statistical problem of estimating peaks in a set of data. This is a problem in shape constrained non-parametric regression. We discuss it in the next section.

\section{Epidemics and Peaks}
\subsection{Taut string}
Epidemics are peaks in the mortality data. Of interest could be the number of peaks, their beginning and end, their heights and locations, and their power. The same problem occurs in X-Ray diffractograms and was treated in \cite{DAGAMEMEMI08}. The approach given here is a simplified version of that paper. It is based on the so called `taut string' algorithm of  \cite{DAVKOV01}. The expression `taut string' is taken from \cite{BABABRBR72}. 

Three  properties of the taut string are here of interest. Firstly, it is very fast requiring about 0.3 seconds for the England and Wales  daily data from 1970 to 2014 of size 16346. Secondly, it minimizes the number of local extremes subject to multiresolution conditions on the residuals. Thirdly, the version used here minimizes the sum of squared residuals between a local minimum and the next local maximum subject to the function being non-decreasing with the corresponding property between a local maximum and the next local minimum. This is known as isotone regression, see  \cite{BABABRBR72}, but here with the extra property that the minimum and maximum values are given by the taut string. In particular it minimizes the sum of squared residuals subject to the shape constraints satisfied by the taut string. The resulting residuals will give a very good estimate of the random variability of the data. 

\begin{figure}[t]
  \centering
\includegraphics[width=.8\textwidth,height=150px]{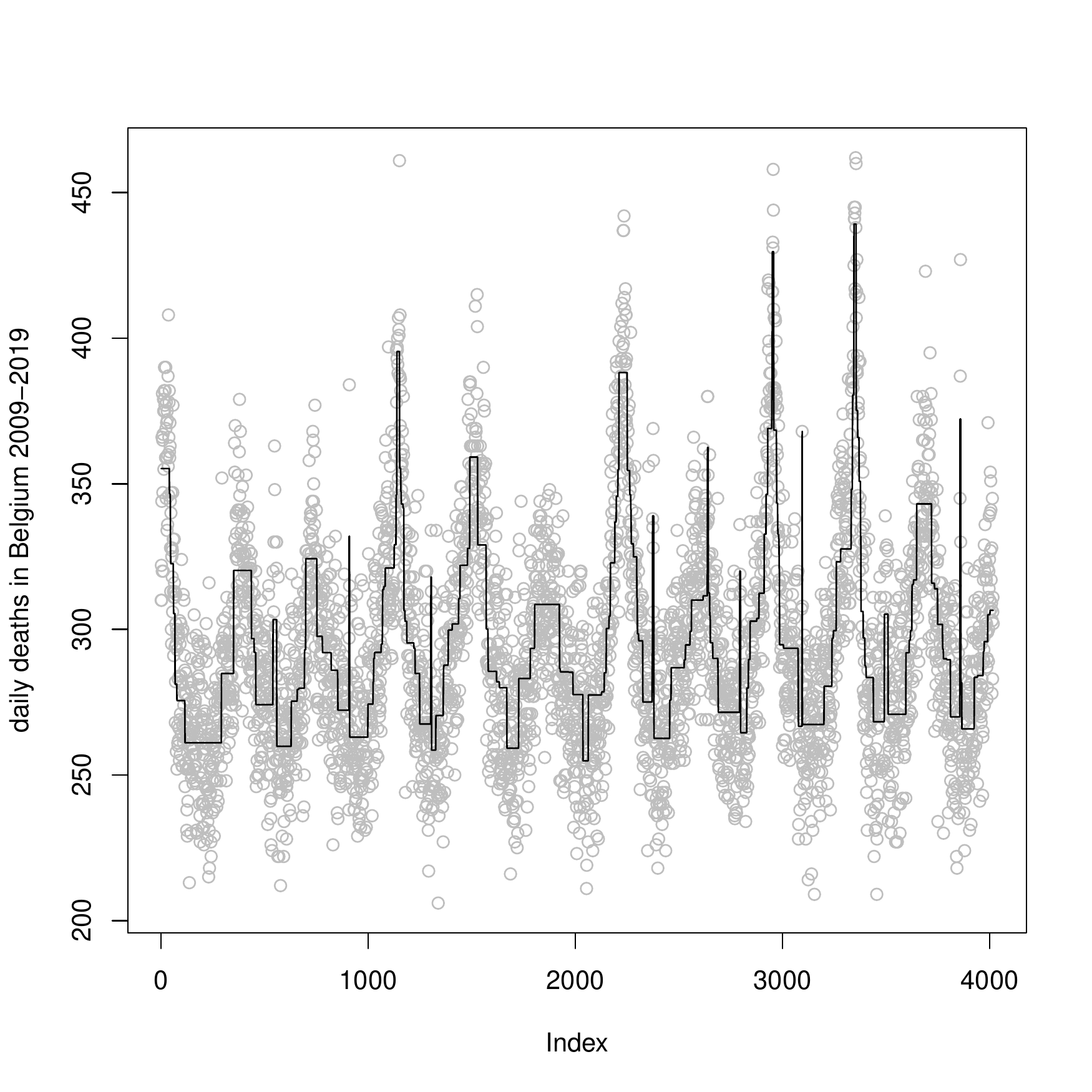}
\includegraphics[width=.8\textwidth,height=150px]{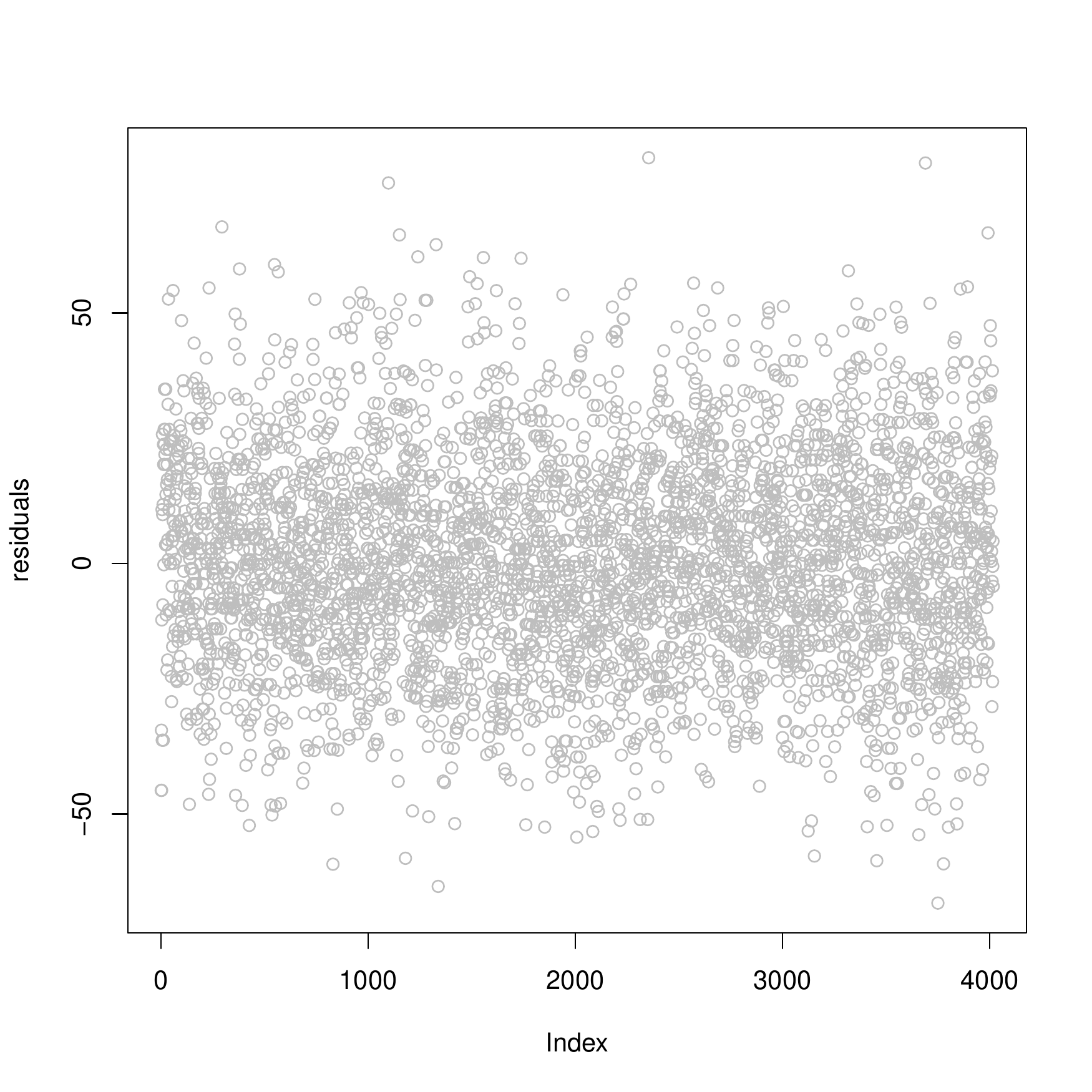}
\includegraphics[width=.8\textwidth,height=150px]{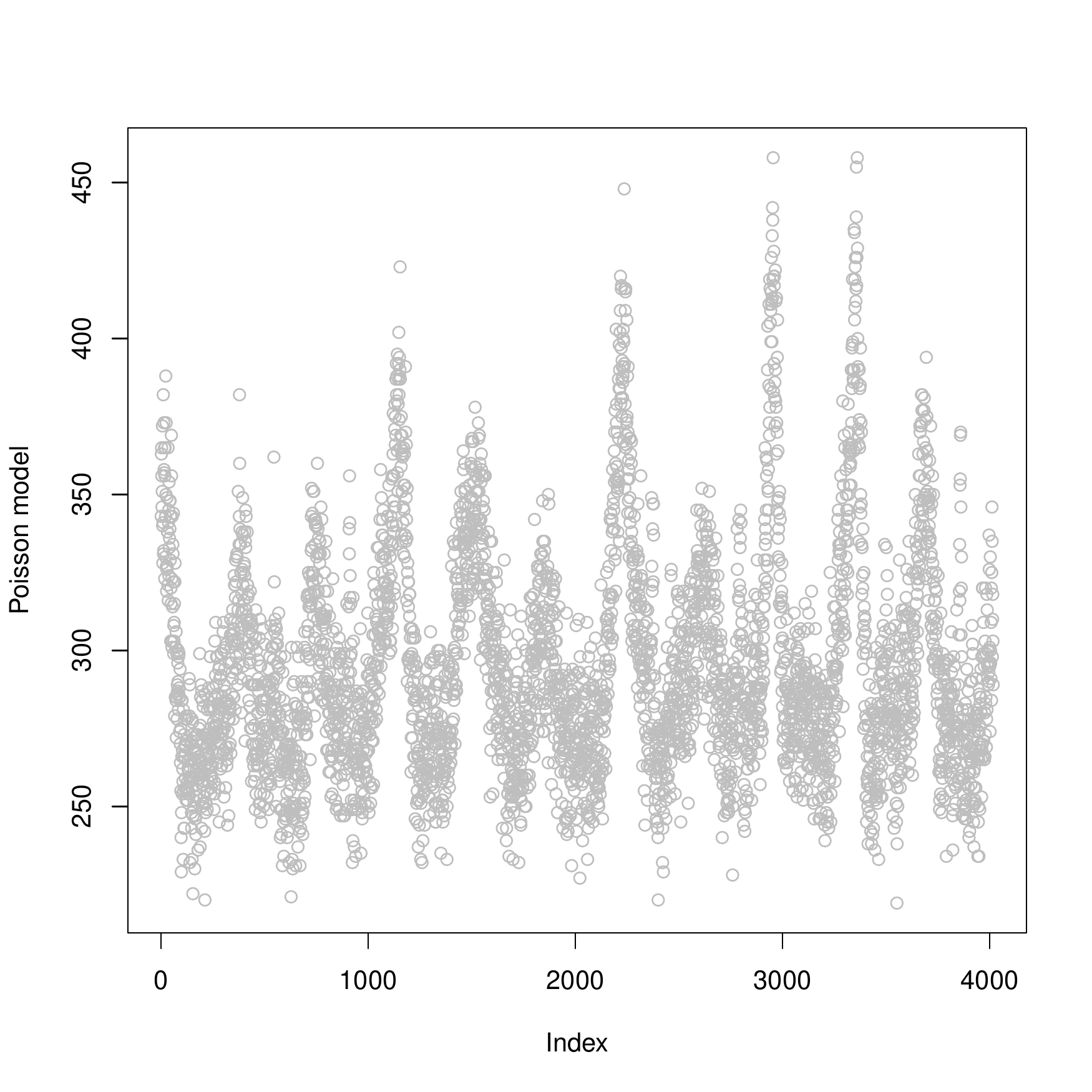}
\caption{Top panel: the Belgian daily data from 2009-2019 together with the taut string reconstruction. Centre panel: the residuals. Bottom panel: a Poisson process based on the taut string.}
\label{fig:belg_daily}
\end{figure}

The top panel of  Figure~\ref{fig:belg_daily}   shows the Belgian daily data from 2009-2019 together with the taut string regressing function. The centre panel shows the residuals and the bottom panel a Poisson process with the taut string as the parameter. The taut string is piecewise constant as is isotonic regression. The location of local minima and maxima is taken to be the midpoint of the interval defining this local extreme value. For the Belgian data these are\\

\begin{tabular}{c}
203,  393,  499,  549,  593,  725,  880,  909,  954, 1145, 1275, 1304,1316\\

1509, 1698, 1863, 2049, 2230, 2348, 2375, 2417, 2639, 2741, 2795, 2812\\

2954, 3085, 3095, 3148, 3350,3465, 3500, 3552, 3683, 3833, 3858, 3895\\
\end{tabular} 
\newline\\
starting with a local minimum at day 203. 

It seems more consistent to take the start of an epidemic not as the midpoint but as the right endpoint of a local minimum interval. Similarly the end of an epidemic is the left endpoint. Thus the first influenza epidemic starts at about day 215 with about  262 deaths a day, peaks at day 393 with about 320 deaths per day and finishes on about day 454 with about 274 deaths per day. These values are taken from the taut string. It lasted for about 239 days and the  total number of deaths in this period was 69619. 
 
The residuals shown in the centre panel of Figure~\ref{fig:belg_daily} have a standard deviation of 20.66. The Poisson reconstruction has one of 17.23 giving an over dispersion of 20\% when compared with the Poisson process. This has been noted before (see \cite{ARMU20}) but not quantified.  It is just about visible to the eye when one compares the top and bottom panels of  Figure~\ref{fig:belg_daily}. If one repeats this for the daily data for Germany from 2016-2019 the taut string residuals have a standard deviation of 90.6 whereas the Poisson model based on the taut string one of 54.8. The over dispersion is about 65\%. For England and Wales daily data exists for the years 2011-2014. The over dispersion is calculated as 25\%. 

The residuals look random but may still contain some residual structure not removed by the taut string. To examine this we consider periodicities and lags. The data length is 4017. We now generate sine and cosine functions 
\[\sin(\pi j(1:n)/n),\cos(\pi j(1:n)/n),j=1,\ldots,n/2\]
with $n=4017$, regard these as covariates and use the Gaussian selection procedure (\cite{DAVDUEM20}) to select covariates. A cut-off P-value of 0.01 resulted four functions being selected. Three had periodicities of almost exactly seven days with a smallest P-value 1.1e-61. The fourth had a period of 365 days and a P-value 4.4e-07. In the case of lags we allowed for lags of up to 500 days. With a cut-off P-value of 0.01 10 lags were selected. As was to be expected the lag of order one was dominant with a P-value 1.3e-21. The P-values of the remaining nine all exceeded 2.0e-07.

\subsection{Smoothing the taut string}
\begin{figure}[b]
  \centering
\includegraphics[width=.8\textwidth,height=150px]{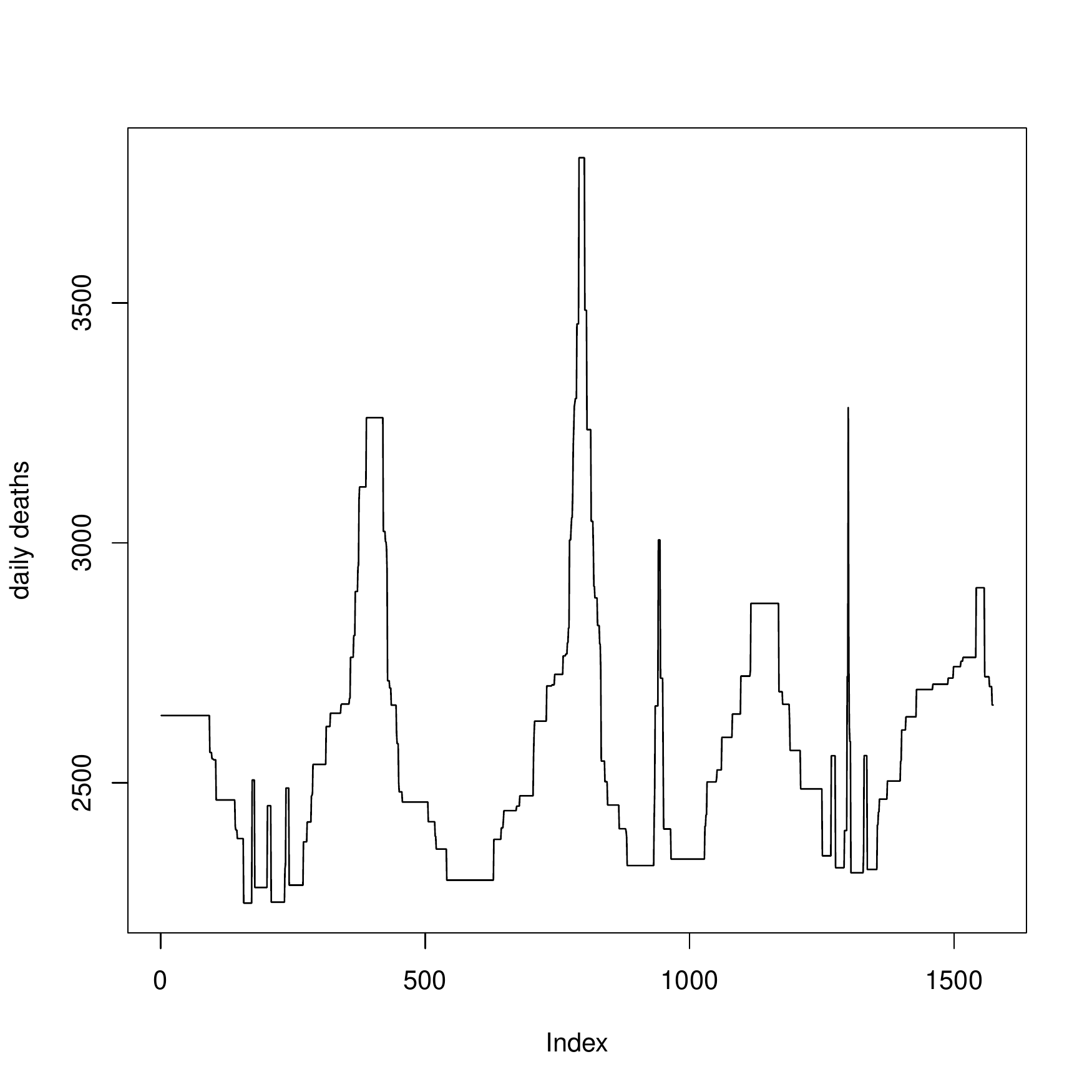}
\includegraphics[width=.8\textwidth,height=150px]{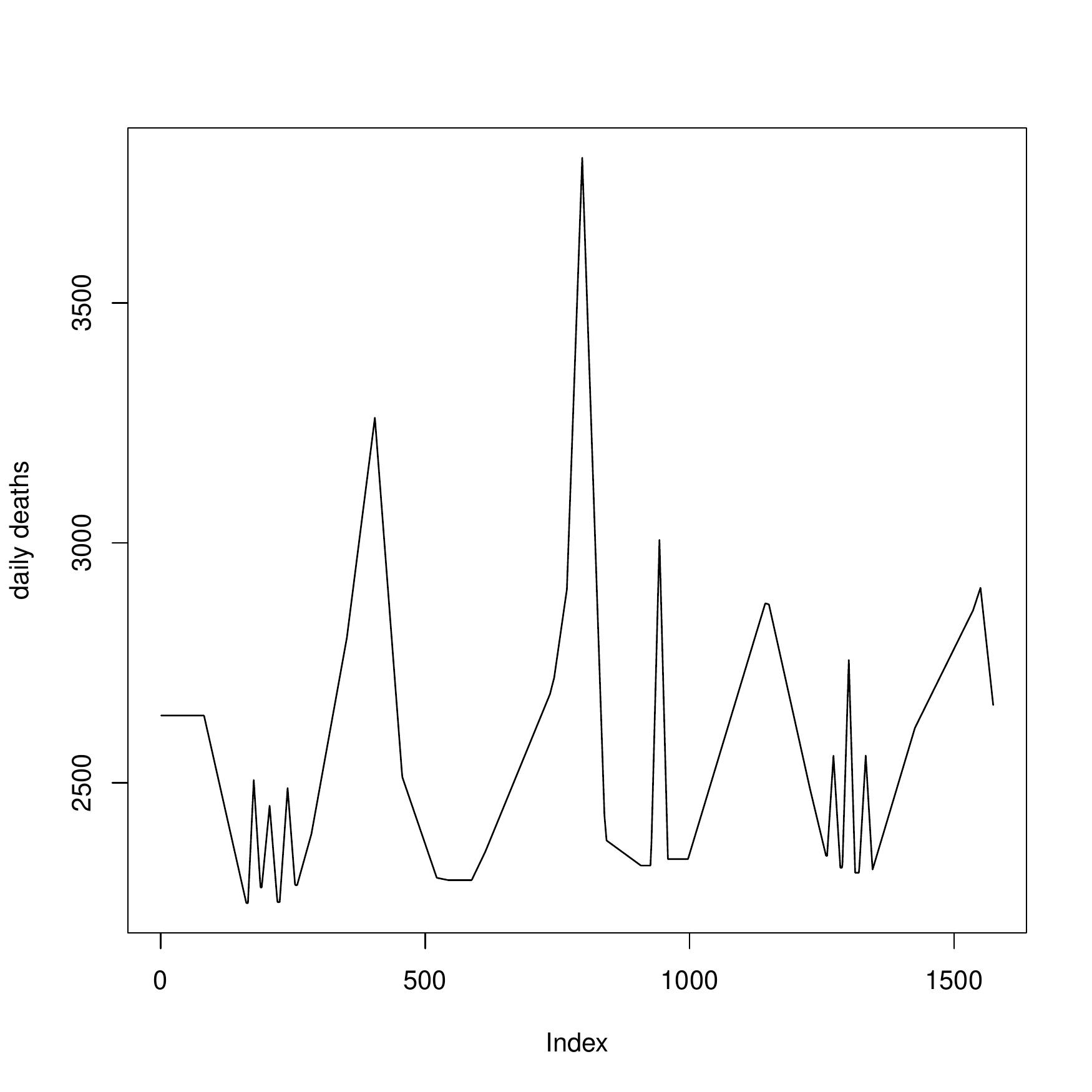}
\includegraphics[width=.8\textwidth,height=150px]{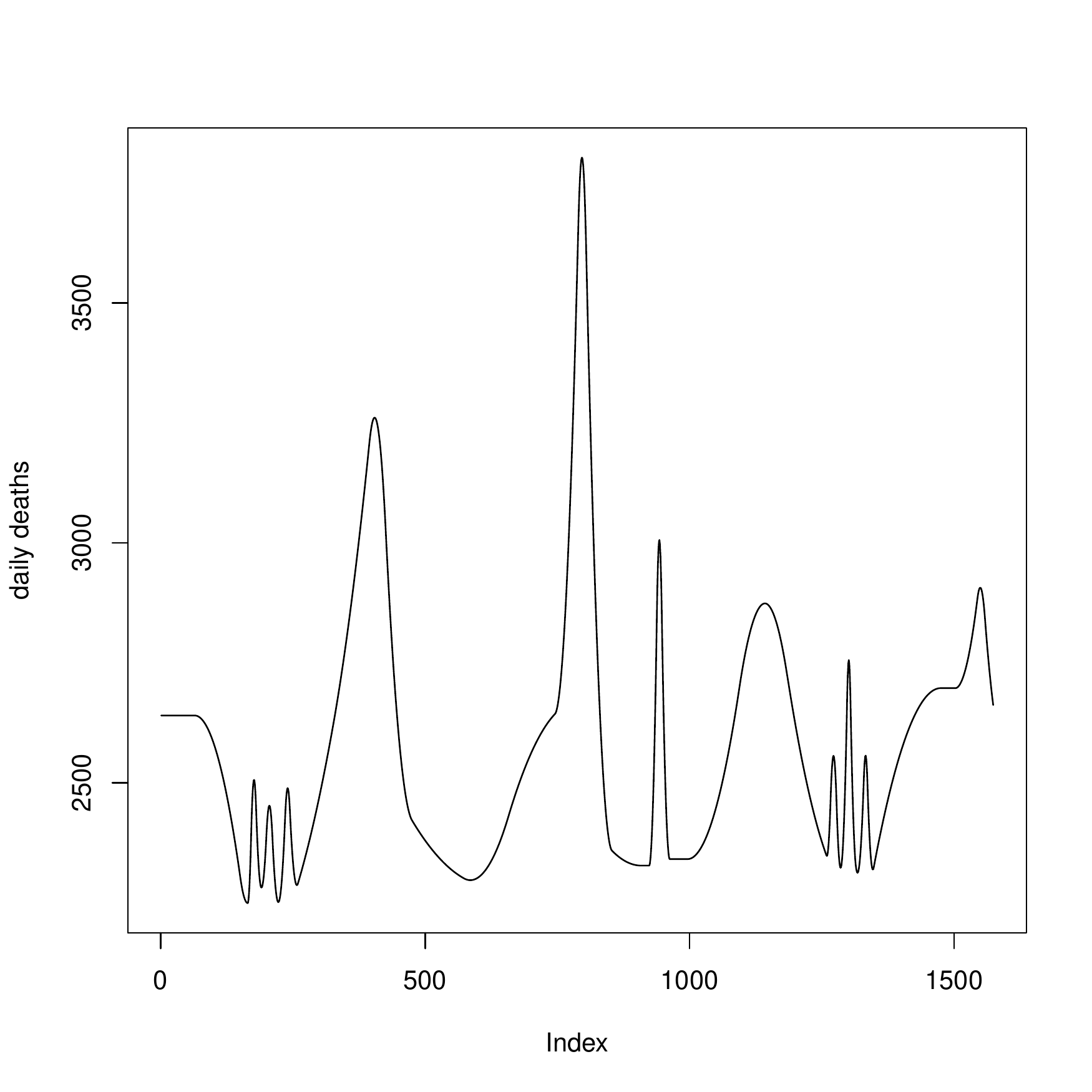}
\caption{Top: the taut string. Centre: minimizing the total variation of the first derivative. Bottom: minimizing the total variation of the second derivative}
\label{fig:tv}
\end{figure}

The main disadvantage of the taut string method is that the resulting function is piecewise constant. To obtain smoother functions, that is one which are differentiable, is considerably more difficult as determining a derivative is an ill-posed problem. Some form of regularization is required. We do this by minimizing the total variation of the derivative subject to the multiscale and the monotonicity constraints of the taut string. This is accomplished here by solving a linear programming problem which causes a large increase in computing time; \cite{KOV07} has a much faster but less stable version. The upper panel of Figure~\ref{fig:tv} shows the result of minimizing the total variation of the first derivative, the lower panel that for the second derivative. The first function is piecewise linear, the second piecewise quadratic.The computing times were about ten minutes. The standard deviations of the residuals were respectively 106 and 106.5, much larger than the standard deviation of the taut string residuals.

\section{ Acknowledgements}
I thank  Lutz D\"umbgen and John Muellbauer for instructive exchanges of emails.

 \bibliographystyle{apalike}
 \bibliography{literature}

\begin{thebibliography}{}

\bibitem[Aron and Muellbauer, 2020]{ARMU20}
Aron, J. and Muellbauer, J. (2020).
\newblock Measuring excess mortality: the case of england during the covid-19
  pandemic.
\newblock Technical Report Oxford Working Paper No. 2020-11, INET.

\bibitem[Barlow et~al., 1972]{BABABRBR72}
Barlow, R.~E., Bartholomew, D.~J., Bremner, J.~M., and Brunk, H.~D. (1972).
\newblock {\em Statistical Inference under Order Restrictions}.
\newblock Wiley, New York.

\bibitem[Davies and D\"umbgen, 2020]{DAVDUEM20}
Davies, L. and D\"umbgen, L. (2020).
\newblock Covariate selection based on a model-free approach to linear
  regression with exact probabilities.
\newblock arxiv.org/abs/1906.01990v2.

\bibitem[Davies et~al., 2008]{DAGAMEMEMI08}
Davies, P.~L., Gather, U., Meise, M., Mergel, D., and Mildenberger, T. (2008).
\newblock Residual based localization and quantification of peaks in x-ray
  diffractograms.
\newblock {\em Annals of Applied Statistics}, 2(3):861--886.

\bibitem[Davies and Kovac, 2001]{DAVKOV01}
Davies, P.~L. and Kovac, A. (2001).
\newblock Local extremes, runs, strings and multiresolution (with discussion).
\newblock {\em Annals of Statistics}, 29(1):1--65.

\bibitem[Farrington et~al., 1996]{FAANBECA96}
Farrington, C., Andrews, N., Beale, A., and Catchpole, M. (1996).
\newblock A statistical algorithm for the early detection of outbreaks of
  infectious disease.
\newblock {\em Journal of the Royal Statistical Society, Series A},
  159:547--563.

\bibitem[Huber, 2011]{HUB11}
Huber, P.~J. (2011).
\newblock {\em Data Analysis}.
\newblock Wiley, New Jersey.

\bibitem[Kovac, 2007]{KOV07}
Kovac, A. (2007).
\newblock Smooth functions and local extreme values.
\newblock {\em Computational Statistics and Data Analysis}, 51(10):5155 --
  5171.

\end{thebibliography}

\begin{figure}[t]
  \centering
\includegraphics[width=.8\textwidth,height=150px]{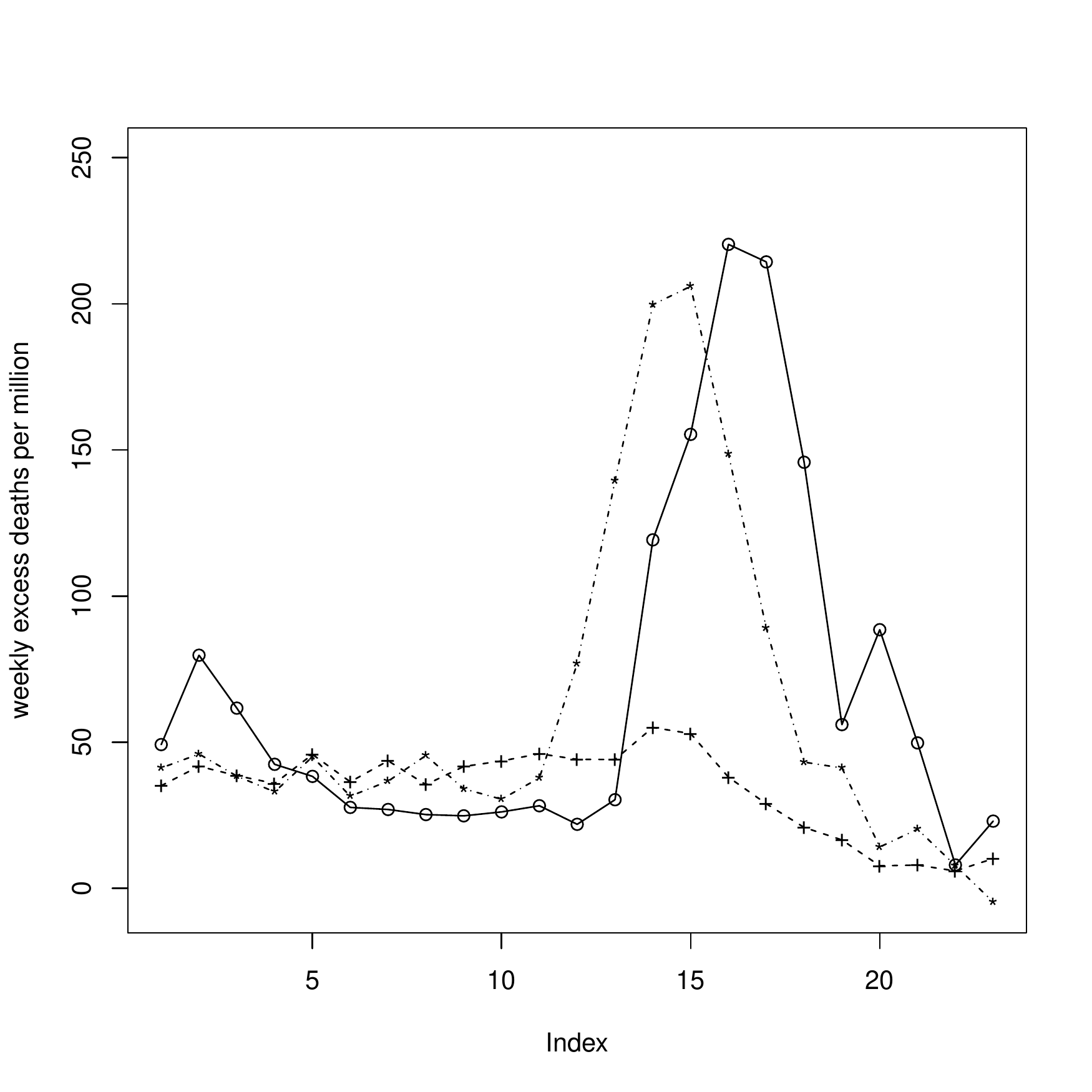}
\includegraphics[width=.8\textwidth,height=150px]{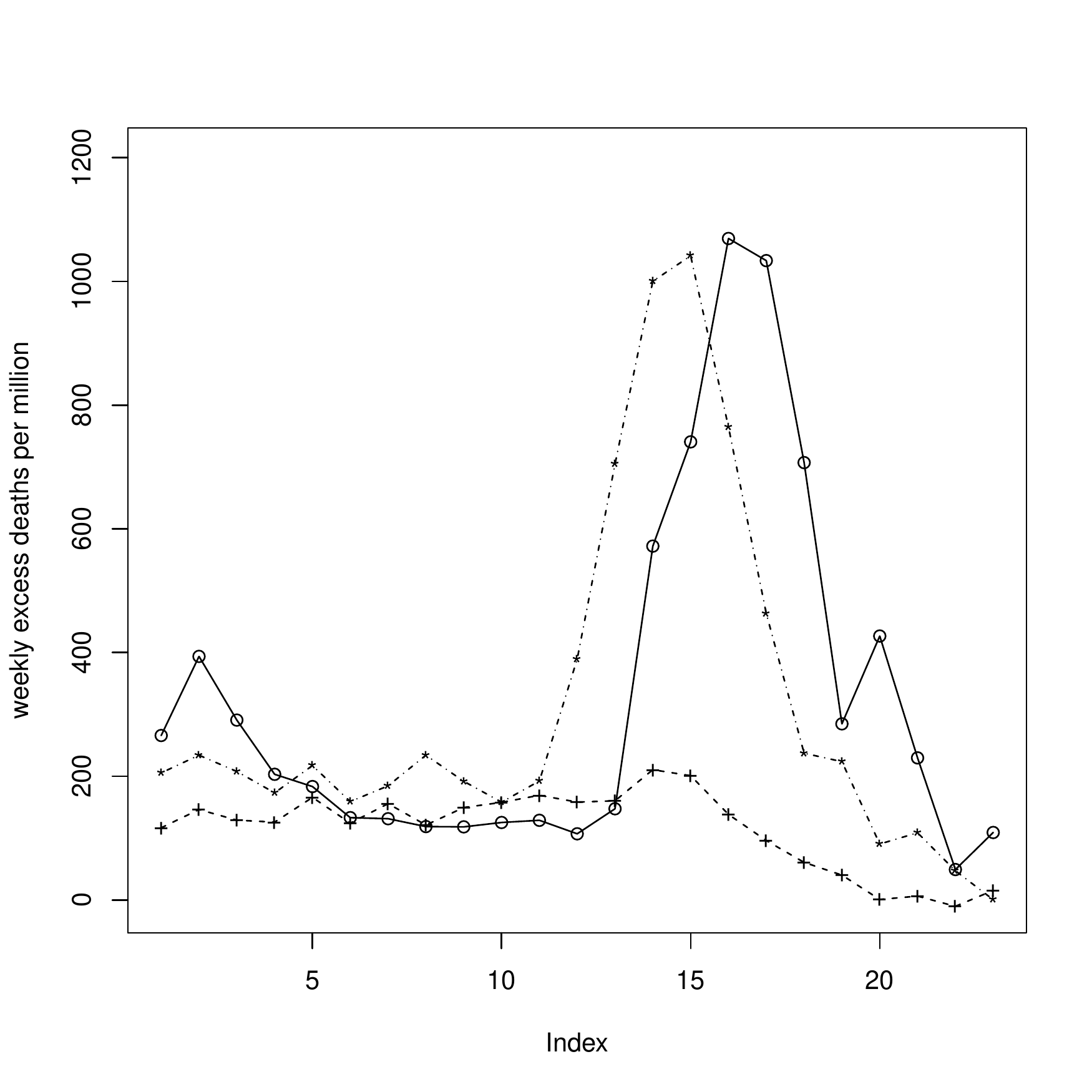}
\includegraphics[width=.8\textwidth,height=150px]{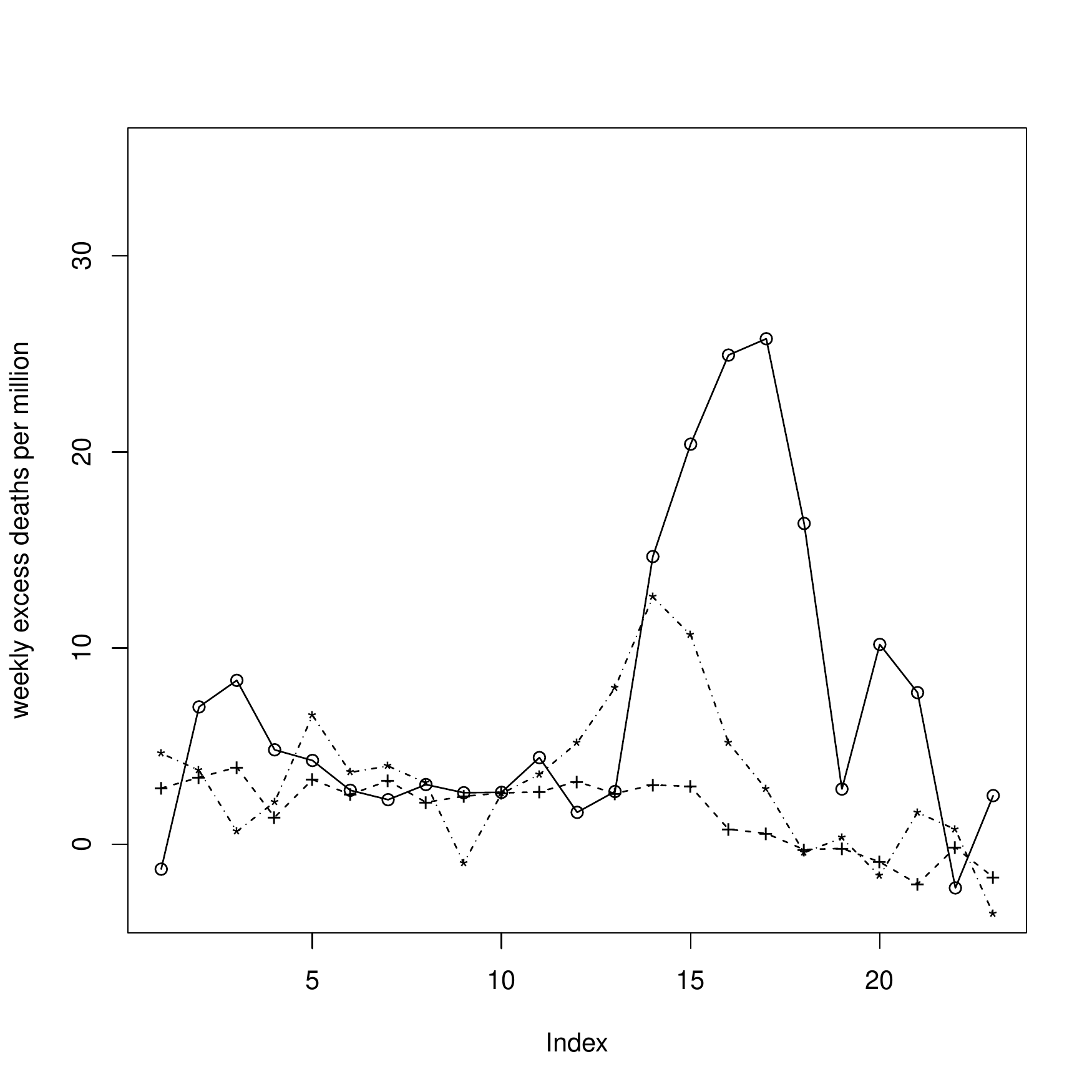}
\caption{The weekly excess deaths for England and Wales (o), Germany (+),  and Belgium (*) for the first 23 weeks of 2020 using the quantile baselines of Table~\ref{tab:10_quantile}: top panel, all ages; centre, over 65; bottom, under 65.}
\label{fig:de_ew_excess_deaths_q}
\end{figure}

\end{document}